\RequirePackage{fix-cm}
\documentclass[twocolumn]{svjour3}          
\smartqed  
\usepackage{graphicx}
\usepackage{graphicx}
   \graphicspath{{../pdf/}{../jpeg/}}
   \DeclareGraphicsExtensions{.pdf,.jpeg,.png}
   
\usepackage{xspace}
\usepackage{url}

\newtheorem{defn}{Definition} 

\newcommand{\etal}{et al.\xspace}
\newcommand{\ie}{i.e.,\xspace}

\newcommand{\fig}[1]{Figure~\ref{#1}}
\newcommand{\tab}[1]{Table~\ref{#1}}
\newcommand{\sect}[1]{Section~\ref{#1}}

\newcommand{\cran}{{CRAN}\xspace}
\newcommand{\rubygems}{{RubyGems}\xspace}
\newcommand{\npm}{{npm}\xspace}
\newcommand{\github}{{GitHub}\xspace} 

\newcommand{\ruby}{{Ruby}\xspace}
\newcommand{\R}{{R}\xspace}
\newcommand{\javascript}{{JavaScript}\xspace}

\begin{document}

\title{An Empirical Comparison of Developer Retention\\
in the RubyGems and npm Software Ecosystems}
\author{Eleni Constantinou        \and
        Tom Mens
}

\institute{E. Constantinou \at
              Software Engineering Lab,\\
	COMPLEXYS Research Institute\\
	University of Mons\\
	Mons, 7000, Belgium\\
              \email{eleni.constantinou@umons.ac.be}
           \and
           T. Mens \at
              Software Engineering Lab,\\
	COMPLEXYS Research Institute\\
	University of Mons\\
	Mons, 7000, Belgium\\
	 \email{tom.mens@umons.ac.be}
}

\date{Received: date / Accepted: date}

\maketitle

\begin{abstract}
Software ecosystems can be viewed as socio-technical networks consisting of technical components (software packages) and social components (communities of developers) that maintain the technical components. Ecosystems evolve over time through socio-technical changes that may greatly impact the ecosystem's sustainability. Social changes like developer turn\-over may lead to technical degradation. This motivates the need to identify those factors leading to developer abandonment, in order to automate the process of identifying developers with high abandonment risk. This paper compares such factors for two software package ecosystems, RubyGems and npm. We analyse the evolution of their packages hosted on GitHub, considering development activity in terms of commits, and social interaction with other developers in terms of comments associated to commits, issues or pull requests. We analyse this socio-technical activity for more than 30k and 60k developers for RubyGems and npm respectively. We use survival analysis to identify which factors coincide with a lower survival probability. Our results reveal that developers with a higher probability to abandon an ecosystem: do not engage in discussions with other developers; do not have strong social and technical activity intensity; communicate or commit less frequently; and do not participate to both technical and social activities for long periods of time. Such observations could be used to automate the identification of developers with a high probability of abandoning the ecosystem and, as such, reduce the risks associated to knowledge loss.

\end{abstract}

\section{Introduction}
\label{sec:introduction}

In the past, software was mainly developed as part of individual and isolated projects~\cite{Blincoe:2015}, while nowadays software projects become more and more interdependent, forming large ecosystems. Such \emph{software ecosystems} are defined by Lungu as ``collections of software projects that are developed and evolve together in the same environment"~\cite{Lungu2008}. We adhere to this definition, and view software ecosystems as \emph{socio-technical networks} comprising a combination of technical components (e.g., software package dependency networks and their source code history) and social components (e.g., communities of contributors involved in the development and maintenance of the software). Well-known examples of software ecosystems are distributions of Linux operating systems and package managers for specific programming languages such as \cran for \R, \rubygems for \ruby and \npm for \javascript. 
Projects that are part of a software ecosystem differ from isolated software projects in the sense that they share source code (e.g., by depending on shared libraries) and developers~\cite{Scacchi:2007}.

While the research community has thoroughly studied individual project evolution, the evolution of software ecosystems is still an emerging research topic~\cite{Serebrenik2015}. 
In particular, there is a need to further investigate social aspects and more precisely, how developers interact in order to keep their projects up to date as part of a sustainable software ecosystem.
Determining the impact of social characteristics of the developer community, and changes in these characteristics over time is an active topic of study~\cite{Foucault:2015,Vasilescu:2015}. In a healthy and sustainable ecosystem, developers can contribute to multiple projects and migrate to other projects within the ecosystem, while problems arise when developers abandon the ecosystem altogether~\cite{Wahyudin:2007}. Considering that developers are the backbone of successful ecosystem evolution, it is important to identify factors that affect developer retention. This knowledge can be used to assist in the future evolution of an ecosystem by predicting possible abandonment risks and mitigating them early. This will increase retention of important developers and reduce the negative effect of such developers leaving the community~\cite{Constantinou:2017}.

This paper aims to identify factors that indicate developers with a high probability of abandoning an ecosystem. In particular, we study the socio-technical evolution of two long-lived software ecosystems, \npm and \rubygems, which correspond to the package managers of the  \javascript and \ruby programming language, respectively. We consider their subset of packages that are developed on \github since we investigate both the social (communication) and technical (coding) activity of software developers involved in these ecosystems.
The coding activity is extracted from source code commits, while the social activity (communication between developers) is based on \github commenting mechanisms (commit, issue and pull request comments) that involve multiple developers.

For each studied ecosystem, we use the statistical technique of \emph{survival analysis} to assess the effect of different types of socio-technical activities in the ecosystem. More precisely, we explore which factors are indicative of developers with increased probability of abandoning the ecosystem.
Such information could assist in automating the process of predicting developers who are likely to abandon the ecosystem. Project managers could use this information to try to motivate these developers to remain active, or to identify developers that are able to take over the abandoners' activities.

The remainder of the paper is structured as follows: Section~\ref{sec:related_work} discusses related work and  Section~\ref{sec:research_hypotheses} formulates our research hypotheses. Section~\ref{sec:experimental_setup} describes the experimental setup, including the data collection method and the operationalisation of our research hypotheses. Section~\ref{sec:results} presents the experimental results and Section~\ref{sec:resultscoreperipheral} reports our findings for core and peripheral developers. Section~\ref{sec:discussion} discusses the limitations of our study and Section~\ref{sec:threats_to_validity} reports implications and threats to the validity of our work. Finally, Section~\ref{sec:conclusion} concludes and provides directions of future research.

\section{Related Work}
\label{sec:related_work}

\subsection{Developer Retention}
Developer retention has been investigated by the research community, including studies aiming to comprehend the factors behind developers abandoning open source software projects~\cite{Ehls:2017} and measuring developer turnover~\cite{Vasilescu:2015,Foucault:2015,Ferreira:2017,Constantinou:2017}, i.e., measuring the size of changes due to social modifications like developers leaving or joining projects.

In a recent work, Lin et al.~\cite{Lin:2017} studied five industrial open source projects in order to identify how developer turnover is affected by the duration and types of contributions. They applied survival analysis to examine the impact of four factors on the duration of developer contributions. Their results show that developers who started contributing earlier stay longer, developers who maintain their own files stay for shorter periods of time, developers who mainly modify files remain in the project longer than those who create files, and developers who mainly code stay longer. As an extension of this work, we focus on both the \emph{social} and \emph{technical} activity, as well as the \emph{frequency} and \emph{intensity} of each type of activity. 

Au{\'e} et al.~\cite{Aue:2016}  investigated software project growth with respect to social diversity. They measured project growth in terms of the number of commits, team members, pull requests and comments. Their results show a statistically weak correlation between project success and diversity in terms of gender and geographical location of contributors. Foucault et al.~\cite{Foucault:2015} characterised patterns of internal and external turnover to indicate the developer mobility inside and outside a project, respectively. They used these patterns to observe a negative effect of external turnover on software quality, based on an analysis on five open source systems. Vasilescu et al.~\cite{Vasilescu:2015} investigated the relationship between gender and tenure diversity, and found that diversity is positively correlated with productivity. 

Constantinou and Mens investigated the socio-technical evolution of Ruby on Rails~\cite{Constantinou:2016} and the entire \ruby ecosystem in GitHub~\cite{Constantinou:2017}. They investigated the effect of permanent changes on the social and technical aspects of evolution of the ecosystem. For each aspect they measured the permanently changed entities with respect to new and obsolete entities and found that large social changes impact the technical evolution of the ecosystem.
In the current follow-up article, we consider both \rubygems and \npm, and also investigate the factors affecting developer retention, using the statistical technique of survival analysis.

Yamashita \etal~\cite{Yamashita2014MSR} analyzed migration trends in a collection of GitHub projects to measure their magnetism (the ability to attract newcomers) and their stickiness (the ability to retain existing developers).
They found that sticky projects are more frequent than magnet projects.
They also analyzed project evolution over time and used quadrant plots to identify projects that are at risk of becoming obsolete. While our approach is related, we analyze developer contributions at the ecosystem level. This implies that developers may contribute to multiple projects, and stay in the ecosystem even though they may abandon some projects or start working on new projects.

Zhou and Mockus~\cite{Zhou2012ICSE} quantified contributors' willingness and their environment to model the chances of becoming a valuable contributor to a project. They used issue tracker data of Mozilla and Gnome to analyse contributor activity and found that joiners who comment on issues (instead of reporting new issues) or get at least one reported issue to be fixed have higher chances of becoming long term contributors. They also found that a productive and clustered peer group increases the odds of becoming long term contributors compared to high popularity and low attention from peers. In contrast to this work, we focus on development activity (in terms of commits) instead of issue tracker data.

Several studies focus on ways to increase the retention of newcomers to software projects~\cite{Steinmacher:2013,Steinmacher2014,Steinmacher:2015}. Steinmacher et al.~\cite{Steinmacher:2015} conducted a systematic literature review to identify and classify the barriers that newcomers face. Their study proposes a model composed of five categories, namely social interactions, previous knowledge, finding a way to start, documentation and technical hurdles. They found that the most evidenced barriers are the lack of social interaction with project members, not receiving (timely) answers and previous technical experience. Also, they highlight the lack of evidence of a causal relationship between the social interaction issues and newcomer success. The current article does not focus on newcomers since our goal is to recover factors affecting developers retention based on their past activity in the ecosystem.

\subsection{Survival Analysis}
Survival analysis \cite{Kleinbaum2012} has been used in a variety of scientific domains (such as biomedical and social sciences) where it is used to study factors affecting the time until an event happens for a variety of measurable events (such as child birth, recovering from a disease, switching employment, marriage or divorce).
Survival analysis models estimate the survival rate of a population over time, considering the notion of censoring. Censoring deals with the fact that some elements of the population may leave the study, and that for some other elements the event of interest does not occur during the observation period. So-called Kaplan-Meier estimations are used to produce and compare survival functions. 

Samoladas \etal~\cite{Samoladas2010} applied survival analysis to assess the expected duration of open source software projects. After partitioning projects by type or domain, they compared the survival function of projects across different domains. They observed that survivability increases as projects grow larger.

Decan \etal \cite{Decan2017interaction} used survival analysis to study the survival of particular database-related libraries in software projects as well as to compare the delay between consecutive software package updates in different software packaging ecosystems \cite{Decan2017}. In contrast to these works, we use survival analysis to study retention of developers rather than software packages.

\section{Research Hypotheses}
\label{sec:research_hypotheses}

In order to determine the factors affecting developer retention, we focus on the activity of project developers in both the technical (\ie commits) and the social (\ie communication through GitHub comments) part of the considered ecosystems. 
We consider someone to be a \emph{developer} if (s)he has committed changes to the project's source code repository, regardless of the type of change. This implies that not all commits necessarily include changes to source code files.

We formulate two sets of research hypotheses concerning the social and technical activities of individual developers.
Each set consists of specific hypotheses about the effect of specific characteristics on developer retention.
The remainder of this section formulates and discusses the rationale behind each hypothesis.
The precise operationalisation of each hypothesis will be explained in \sect{sec:experimental_setup}.

\subsection{Research Hypotheses Related to Social Activity}

As mentioned before, we measure social activity of developers in terms of communication between developers through \github comments that are attached to either project commits, pull requests, or issues in the associated issue tracker. To avoid overestimating the social activity, we ignore messages in communication threads that involve only a single developer. In other words, if a developer adds a comment to which no other developer reacts, we do not consider this as a communication activity.

\noindent \textbf{$H_{1.0}$ Developers that do not communicate have a higher probability of abandoning the ecosystem sooner.}
This hypothesis considers the characteristic of whether a developer actually discusses (through GitHub comments) with other developers in the ecosystem.
We distinguish three categories of developers: (1) \emph{socially active}; (2) \emph{socially inactive}; and (3) \emph{social abandoner}.

\noindent \textbf{$H_{1.1}$ Developers that communicate less intensively have a higher probability of abandoning the ecosystem sooner.}
This hypothesis explores the communication intensity in terms of the number of comments contributed by each developer.
We consider four categories of developers in function of their communication intensity, which can be: (1) \emph{very weak}; (2) \emph{weak}; (3) \emph{strong}; or (4) \emph{very strong}.

\noindent \textbf{$H_{1.2}$ Developers that communicate less frequently have a higher probability of abandoning the ecosystem sooner.}
This hypothesis explores the frequency of social activity, i.e., the percentage of active time units during which a developer was socially active, with respect to his entire timespan of socio-technical activity in the ecosystem. 
We classify developers into four categories based on their percentage of socially active periods of time: 
(1) \emph{very rarely}; (2) \emph{rarely}; (3) \emph{frequently}; (4) \emph{very frequently}.

\noindent \textbf{$H_{1.3}$ Developers that do not communicate for a longer period have a higher probability of abandoning the ecosystem sooner.}
This hypothesis focuses on the length of inactive periods of time concerning developers' social activity. 
We measure the length of the largest inactive period of time and compare it against the duration of the entire socio-technical activity.
We classify developers into four categories depending on their longest period of social inactivity: 
(1) \emph{very short}; (2) \emph{short}; (3) \emph{long}; and (4) \emph{very long}.

\subsection{Research Hypotheses Related to Technical Activity}

The hypotheses regarding technical activity (measured in terms of commits) are very similar to those for social activity, except that there is no counterpart for $H_{1.0}$. In order to be considered a developer, one needs to have made commits in the ecosystem for over one month. Considering a category of \emph{technically inactive} developers would be meaningless.

\noindent \textbf{$H_{2.1}$ Developers  that commit less intensively have a higher probability of abandoning the ecosystem sooner.}
This hypothesis focuses on the intensity of the commit activity of a developer.
As for $H_{1.1}$ we again consider four categories of developers in function of their commit intensity, which can be: (1) \emph{very weak}; (2) \emph{weak}; (3) \emph{strong}; or (4) \emph{very strong}.

\noindent \textbf{$H_{2.2}$ Developers that commit  less frequently have a higher probability of abandoning the ecosystem sooner.}
This hypothesis explores the frequency of technical activity, i.e., the percentage of time units that a developer was active in terms of commits, with respect to the timespan of his socio-technical activity in the ecosystem.
We classify developers into four categories based on their percentage of technically active periods of time: 
(1) \emph{very rarely}; (2) \emph{rarely}; (3) \emph{frequently}; (4) \emph{very frequently}.

\noindent \textbf{$H_{2.3}$ Developers that do not commit for longer periods have a higher probability of abandoning the ecosystem sooner.}
This hypothesis focuses on the length of inactive periods of time concerning developers' commit activity. 
We measure the length of the largest inactive period of time and compare it against the duration of the entire technical activity.
We classify developers into four categories depending on their longest period of technical inactivity: 
(1) \emph{very short}; (2) \emph{short}; (3) \emph{long}; and (4) \emph{very long}.

\section{Experimental Setup}
\label{sec:experimental_setup}
\subsection{Data Sources}
To verify our research hypotheses, we use two data sources for each ecosystem: information extracted from the package management system, and development and communication information extracted from GitHub. Initially, we parse all the packages from each ecosystem\footnote{\url{https://rubygems.org/} for RubyGems}\footnote{\url{https://www.npmjs.com/} for npm} in order to acquire the package names, versions, dependencies and information about the repositories hosting their development. Next, we use the GHTorrent dataset~\cite{Gousios13}\footnote{We use the 2016-09-05 dump of the GHTorrent dataset} to acquire the development activity of each package that is hosted on GitHub. In order to match a package with a GitHub repository, we parsed all the available links provided by each package in their information (homepage, bug tracker URI and source code repository link) to recover the ones that are linked to a GitHub repository.

\begin{table}[!t]
\renewcommand{\arraystretch}{1.3}
\caption{Descriptive Statistics of Dataset}
\label{tab:dataset_statistics}
\centering
\begin{tabular}{p{3.9cm}||r|r}
number of \ldots & RubyGems & npm\\
\hline
Packages &  121,960 & 316,453 \\
Packages hosted in GitHub & 69,941 & 178,879\\
& (57,3\%) & (56,5\%)\\
\hline
Developers (commiters) & 56,793 & 119,114 \\
Developers with $>$1 month activity & 31,347 & 63,357 \\
Socially active developers & 22,148 & 44,244\\
\hline
GitHub messages & 1,548,816 & 4,187,235 \\
Git commits & 2,847,398 & 7,769,680\\
\end{tabular}
\end{table}

We found an overlap of 110 packages between the npm and RubyGems ecosystems, \ie packages pointing to the same GitHub repository.
Considering that among these packages some are very popular and intensively developed (e.g., Rails, Selenium, RethinkDB), it is important to assign each package to a single ecosystem so as to eliminate any bias when comparing the results of further analyses between the two ecosystems. To achieve this, we relied on two heuristics: (1) the number of Ruby (.rb) and JavaScript (.js) source code files of each repository; and (2) the number of commits touching Ruby and JavaScript source code files, respectively. For our first heuristic, if the number of files for one programming language is three times higher than for the other, we assign the package to the respective packaging ecosystem. If this heuristic fails to assign the package to an ecosystem, we apply the second heuristic: if source code files of one programming language are touched significantly more times than the other one (three times more commits), then the repository is assigned to the respective ecosystem. 

To verify our research hypotheses we rely on socio-technical data extracted from GHTorrent. As technical data we gather, for each project in the ecosystem, the GitHub \emph{commits}, including commit author and commit date. As social activity data we extract the developers and timestamps of \emph{messages} posted through commenting mechanisms in GitHub commits, issues and pull requests. We exclude all messages in discussions that do not involve any interaction with other developers, \ie where the number of participants to the discussion was equal to 1.

We also removed from our dataset one GitHub account that corresponds to a bot, which we identified due to its unusually large number of commits (163,166) and confirmed by its account name \texttt{greenkeeperio-bot}. This bot is used by the Greenkeeper tool for automatically updating npm dependencies.

Open source project developers can become temporarily inactive in an ecosystem. In order to determine the (in)active time periods, we use one-month time units. Choosing larger time units (e.g., three or six months) would pose limitations to identifying the frequency of contributions. Firstly, it would suggest that a developer that commits once every three or six months is very frequently active. Secondly, in order to use our approach for developer retention in real-world settings, dayflies and short-lived developers should be excluded. Considering that it is expected that newcomers abandon open source software projects~\cite{Steinmacher:2013}, our main interest lies in identifying why developers that are active longer abandon the ecosystem.

Our dataset considers development data from January 2001 for RubyGems and December 1999 for npm. For both ecosystems, the observation period ends in September 2016. \tab{tab:dataset_statistics} summarises descriptive statistics of the two considered ecosystems, npm and RubyGems. We report the total number of packages and the number of those packages that are hosted on GitHub, and observe that 57.3\% of all RubyGems and  56.5\% of all npm packages are hosted on GitHub.
We also present the number of commit authors in those projects, and the subset of developers having commit activity for more than one month, as well as the subset of developers that are socially active by communicating through GitHub comments with other developers for each ecosystem. The final rows in  \tab{tab:dataset_statistics} show the number of messages registered using the GitHub commenting mechanisms and the number of commits of active developers respectively.

To ensure the reproducibility of our work, a publicly available dataset for each ecosystem and the R scripts we used for this study are available on:\\
\url{http://www.econst.eu/developer-retention-secos}

Next, we explain how we operationalise our research hypotheses.

\subsection{Operationalisation of Social Activity Hypotheses}

\begin{table*}[!t]
\renewcommand{\arraystretch}{1.3}
\caption{Descriptive statistics of developer-based measurements}
\label{tab:measurement_stats}
\centering
\begin{tabular}{p{7cm}||c|c|c||c|c|c}
Measurement & \multicolumn{3}{c||}{RubyGems} & \multicolumn{3}{c}{npm}\\
& Min & Max & Median & Min &Max  & Median\\ 
\hline \hline
Comments & 1 & 16,321 & 13 & 1 & 14,202 & 16 \\ \hline
Months of social activity & 1 & 94 & 6 & 1 & 85 & 6 \\ \hline
Largest timespan of social inactivity (in months) & 2 & 82 & 11 & 2 & 71& 8 \\ \hline
Communication frequency percentage & 1\% & 100\% & 21\% & 1\% & 100\% & 31\% \\ \hline
Social inactivity percentage & 2\% & 100\% & 33\% & 100\% & 100\% & 36\% \\
\hline\hline
Commits & 2 & 8,673 & 24 & 2 & 20,234 & 26 \\ \hline
Months with commits & 2 & 136 & 4 & 2 & 133 & 4 \\ \hline
Largest timespan of commit inactivity (in months) & 2 & 170 &  7 &  2 & 183 &  5 \\ \hline
Commit frequency percentage & 2\% & 100\% & 28\% &  2\% & 100\% & 4\% \\ \hline
\% of months without commit activity & 2\% & 100\% & 67\% &  2\% & 100\%&  67\% \\ 
\end{tabular}
\end{table*}

\noindent \textbf{$H_{1.0}$ Developers that do not communicate have a higher probability of abandoning the ecosystem sooner.} We distinguish three categories based on the presence and timestamp of messages contributed by each developer in discussions with other developers (through GitHub comments). The three categories we use to characterise social activity correspond to:
\begin{itemize}
\item \emph{Socially active:} developers that discussed at least once, and remain doing so.
\item \emph{Socially inactive:} developers that never discussed.
\item \emph{Social abandoner:} developers that initially discussed, but remained inactive in discussions for over one year after their latest message.
\end{itemize}

\noindent \textbf{$H_{1.1}$ Developers that communicate less intensively have a higher probability of abandoning the ecosystem sooner.}
We rely on the number of comments contributed by a developer in discussions to measure his communication intensity. The statistics on socially active developers are presented in the upper half of \tab{tab:measurement_stats}. The distribution of comments is highly skewed, i.e., most of the developers have a small number of messages while only a small portion of the developers are extremely active. We therefore opted to use the median as a central tendency measure (as opposed to using the mean), since it is more robust to outliers and skewed distributions.
Similarly, we use the quartile grouping method to create four bins that classify each developer based on the intensity of social activity~\cite{Lanza2006} (we also report the cut-points for each category and ecosystem):
\begin{itemize}
\item \emph{Very weak:} Developers with a very small number of messages (RubyGems: 1-3, npm: 1-4)
\item \emph{Weak:} Developers with a small number of messages (RubyGems: 4-12, npm: 5-15)
\item \emph{Strong:} Developers with a moderate number of messages (RubyGems: 13-38, npm: 16-50)
\item \emph{Very strong:} Developers with a large number of messages (RubyGems: 39-16,321, npm: 51-14,202)
\end{itemize}

\noindent \textbf{$H_{1.2}$ Developers that communicate less frequently have a higher probability of abandoning the ecosystem sooner.} Communication frequency is defined as the percentage of distinct socially active months with respect to the total duration of the developer's socio-technical activity, i.e., the number of elapsed months from his first till his last contribution. Statistics on communication frequency are presented in the fourth row of \tab{tab:measurement_stats}. We classify each developer based on communication frequency into the following categories: 
\begin{itemize}
\item \emph{Very rarely} active developers have engaged in social activity for a quarter of their overall stay (0-25\%)
\item \emph{Rarely} active developers have engaged in social activity for half of their overall stay (25-50\%)
\item \emph{Frequently} active developers have engaged in social activity for three quarters of their overall stay (50-75\%)
\item \emph{Very frequently} active developers have engaged in social activity throughout their stay (75-100\%)
\end{itemize}

\noindent \textbf{$H_{1.3}$ Developers that do not communicate for a longer period have a higher probability of abandoning the ecosystem sooner.}
Social inactivity is defined as the percentage of the largest number of consecutive months where the developer did not communicate with other developers, with respect to the entire duration of the developer's socio-technical activity. We follow this approach since short-term inactivity periods are expected~\cite{Bosu2014}, especially in the presence of peripheral developers. Similar to the communication frequency, we classify developers based on their percentage of social inactivity into four categories:
\begin{itemize}
\item \emph{Very short} (0-25\%)
\item \emph{Short} (25-50\%)
\item \emph{Long} (50-75\%)
\item \emph{Very long} (75-100\%)
\end{itemize}

\subsection{Operationalisation of Technical Activity Hypotheses}

\noindent \textbf{$H_{2.1}$ Developers  that commit less intensively have a higher probability of abandoning the ecosystem sooner.}
Commit intensity is measured according to the number of commits. Similar to communication intensity, commits also follow a highly skewed distribution (row 8 of \tab{tab:measurement_stats}). The categories and cut-points that stem from quartile grouping are:
\begin{itemize}
\item \emph{Very weak:} Developers with a very small number of commits (RubyGems: 1-7, npm: 1-8)
\item \emph{Weak:} Developers with a small number of commits (RubyGems: 8-23, npm: 9-25)
\item \emph{Strong:} Developers with a moderate number of commits (RubyGems: 24-69, npm: 26-81)
\item \emph{Very strong:} Developers with a large number of commits (RubyGemts: 69-8,763, npm: 82-20,234)
\end{itemize}

\noindent \textbf{$H_{2.2}$ Developers that commit less frequently have a higher probability of abandoning the ecosystem sooner.} Commit frequency is defined as the percentage of distinct months of active commits by the developer with respect to the total duration of the developer's socio-technical activity (rows 9-10 of \tab{tab:measurement_stats}). Each developer is assigned to one of four categories:
\begin{itemize}
\item \emph{Very rarely} active developers have commits for up to a quarter of their overall stay (0-25\%)
\item \emph{Rarely} active developers have commits for up to half of their overall stay (25-50\%)
\item \emph{Frequently} active developers have commits for up to three quarters of their overall stay (50-75\%)
\item \emph{Very frequently} active developers have commits throughout their stay (75-100\%)
\end{itemize}

\noindent \textbf{$H_{2.3}$ Developers that do not commit for longer periods have a higher probability of abandoning the ecosystem sooner.} Commit inactivity is defined as the percentage of the largest number of consecutive months where the developer did not commit with respect to the total duration of the developer's coding activity (rows 11-12 of \tab{tab:measurement_stats}). We classify each developer based on the percentage of commit inactivity where the inactivity periods can be:
\begin{itemize}
\item \emph{Very short} (0-25\%)
\item \emph{Short} (25-50\%)
\item \emph{Long} (50-75\%)
\item \emph{Very long} (75-100\%)
\end{itemize}


\section{Results of Survival Analysis}
\label{sec:results}

Our research hypotheses need to be verified or rejected based on the extracted data.
To understand the effect of each activity type, including its intensity and frequency, on a developer's abandonment in a software ecosystem, we resort to the statistical technique of \emph{survival analysis}~\cite{Kleinbaum2012}.

Survival analysis models estimate the survival rate of a population until the occurrence of an event of interest. 
In this study, the considered population is all developers in the ecosystem, and the event of interest is the abandonment of a developer from the ecosystem. In order to consider a developer as an abandoner, he must have remained inactive for at least one year after his last commit. Given that we have fixed August 2015 as the end of the observation period for our survival analyses, developers having any type of activity after that date are considered as active, while the remaining ones are marked as abandoners.

This section reports our findings for each research hypothesis based on the respective survival analyses. We visualise and discuss our results using Kaplan-Meier survival curves and 95\% confidence intervals (the dotted lines accompanying each survival curve).

We statistically verify each hypothesis using log-rank tests~\cite{Wellek1993} to ensure that the null hypothesis (assuming that all survival curves are the same) can be rejected. For each survival analysis for both ecosystems, we use log-rank tests 
to find significant differences between categories of developers affecting their likelihood of abandoning the ecosystem.
For each considered factor, we carry out a Bonferroni correction to account for the fact that multiple log-rank tests are carried out to test each pair of developer categories.
To obtain a confidence level of 95\%, the significance level used per test is $\alpha = \frac{0.05}{m}$, where $m$ is the number of considered pairs of categories within the same factor. For example, for the factor of communication intensity studied in hypothesis  $H_{1.1}$, there are four different categories of intensity (i.e., very weak, weak, strong and very strong), implying that there are $m=6$ tests (one comparison per pair of categories). Hence, the Bonferroni correction requires to test at a significance level $\alpha=\frac{0.05}{6} = 0.008$. When presenting our results, statistical significance with Bonferroni correction is confirmed unless stated otherwise.

\subsection{Survival Analysis of Social Activity}

\begin{figure}[!ht]
\centering
\begin{tabular}{c}
\includegraphics[scale=0.49]{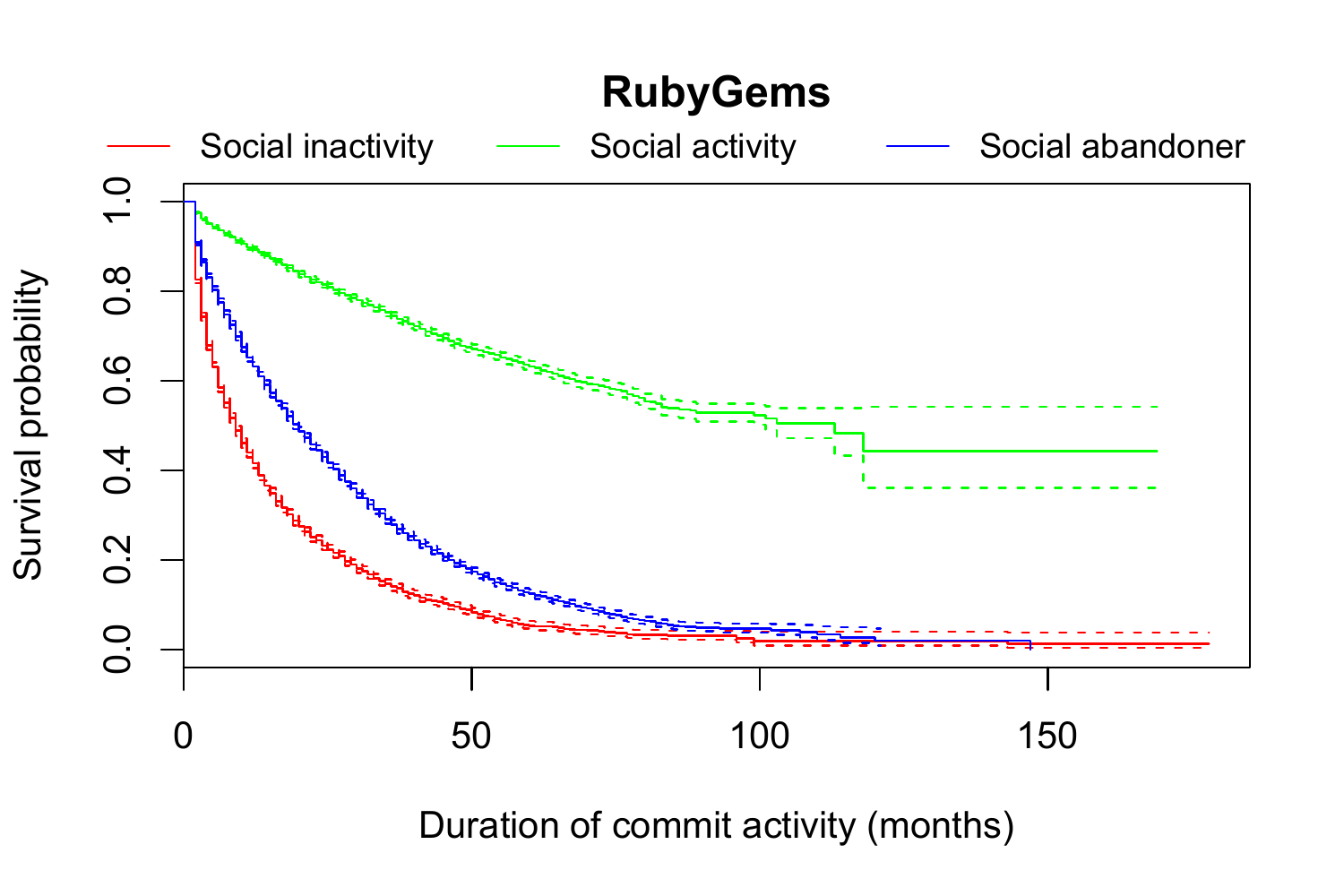}
\\
\includegraphics[scale=0.49]{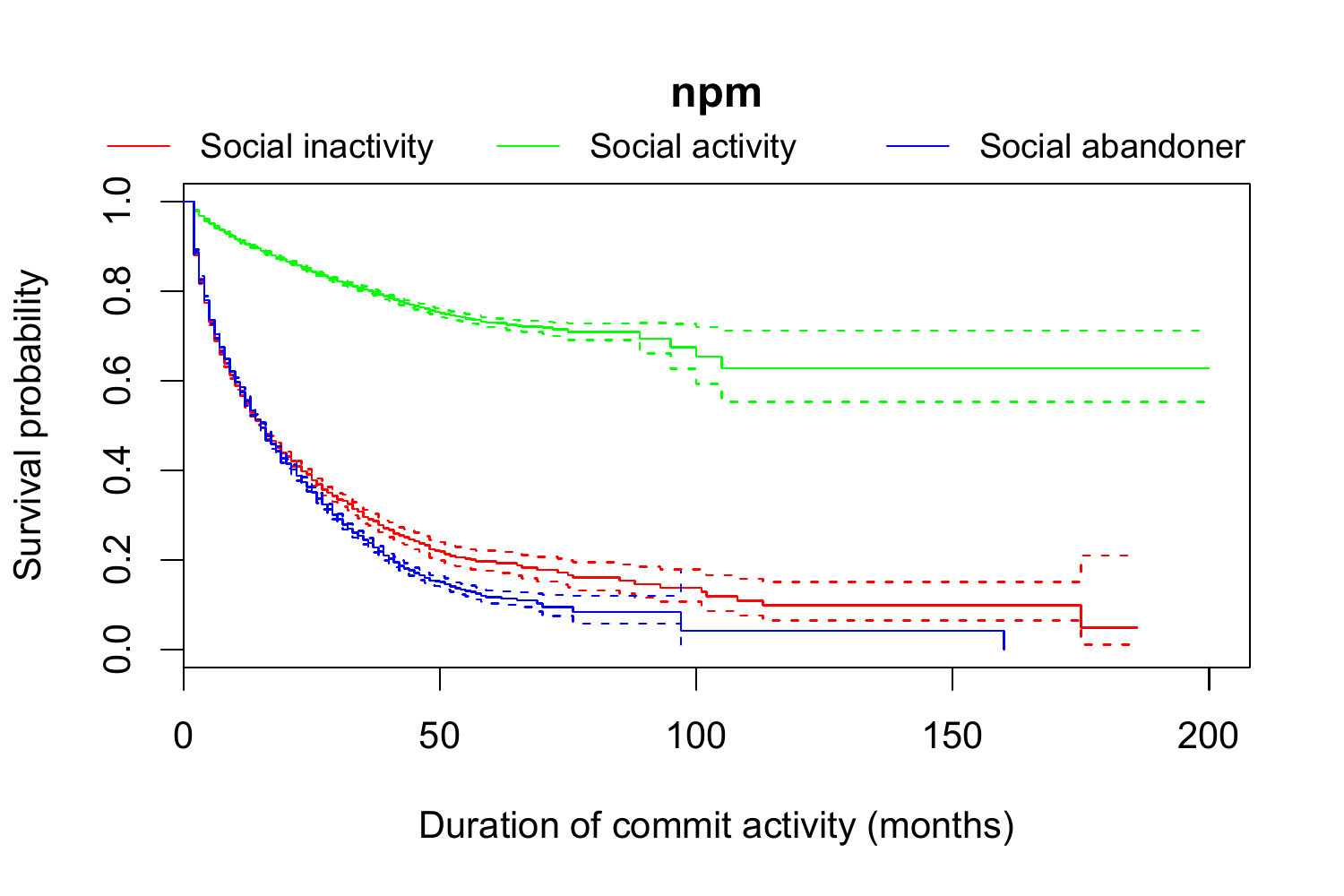}
\end{tabular}
\caption{$H_{1.0}$ -- Survival curves of developers based on social activity}
\label{fig:h1_0}
\end{figure}

\textbf{$H_{1.0}$ Developers that do not communicate have a higher probability of abandoning the ecosystem sooner.}
\fig{fig:h1_0} shows the Kaplan-Meier survival curves for developers belonging to each social activity category for RubyGems and npm.
The survival curve for actively communicating developers is significantly higher than the ones of the other two categories for both ecosystems. Additionally, socially inactive contributors and social abandoners are more likely to abandon the ecosystem from the very beginning of their involvement in the ecosystem, while less than 10\% remain active for over 10 years for both ecosystems. Although the Bonferroni correction did not pass all the tests for the npm ecosystem, there is a statistically significant distinction between the category of socially active contributors and the other two categories. These results confirm our hypothesis that \emph{developers who do not communicate or stopped communicating with other developers at some point are more likely to abandon the ecosystem sooner.}

\subsection*{\textbf{$H_{1.1}$ Developers that communicate less intensively have a higher probability of abandoning the ecosystem sooner.}}
\begin{figure}[!ht]
\centering
\begin{tabular}{c}
\includegraphics[scale=0.49]{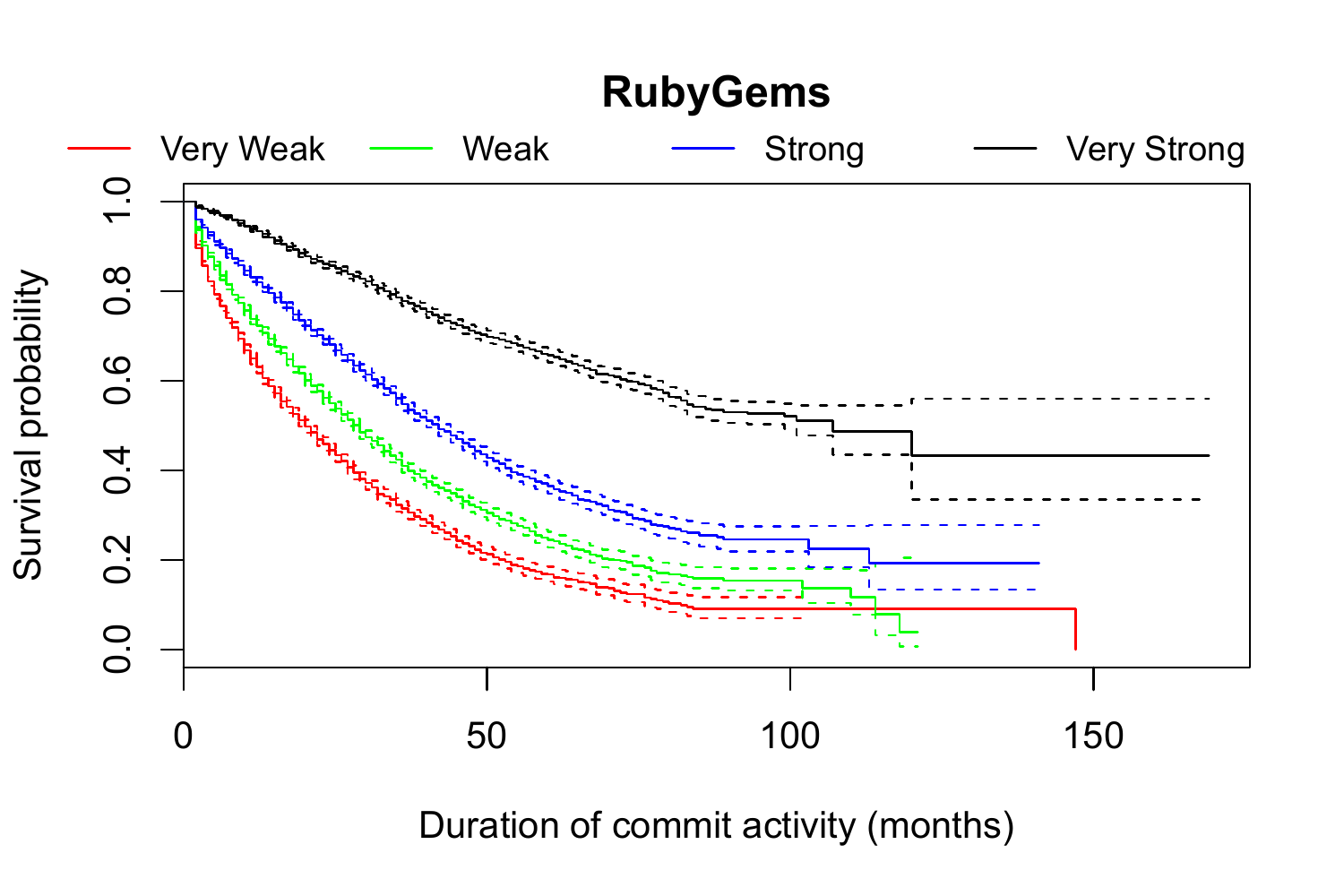}
\\
\includegraphics[scale=0.49]{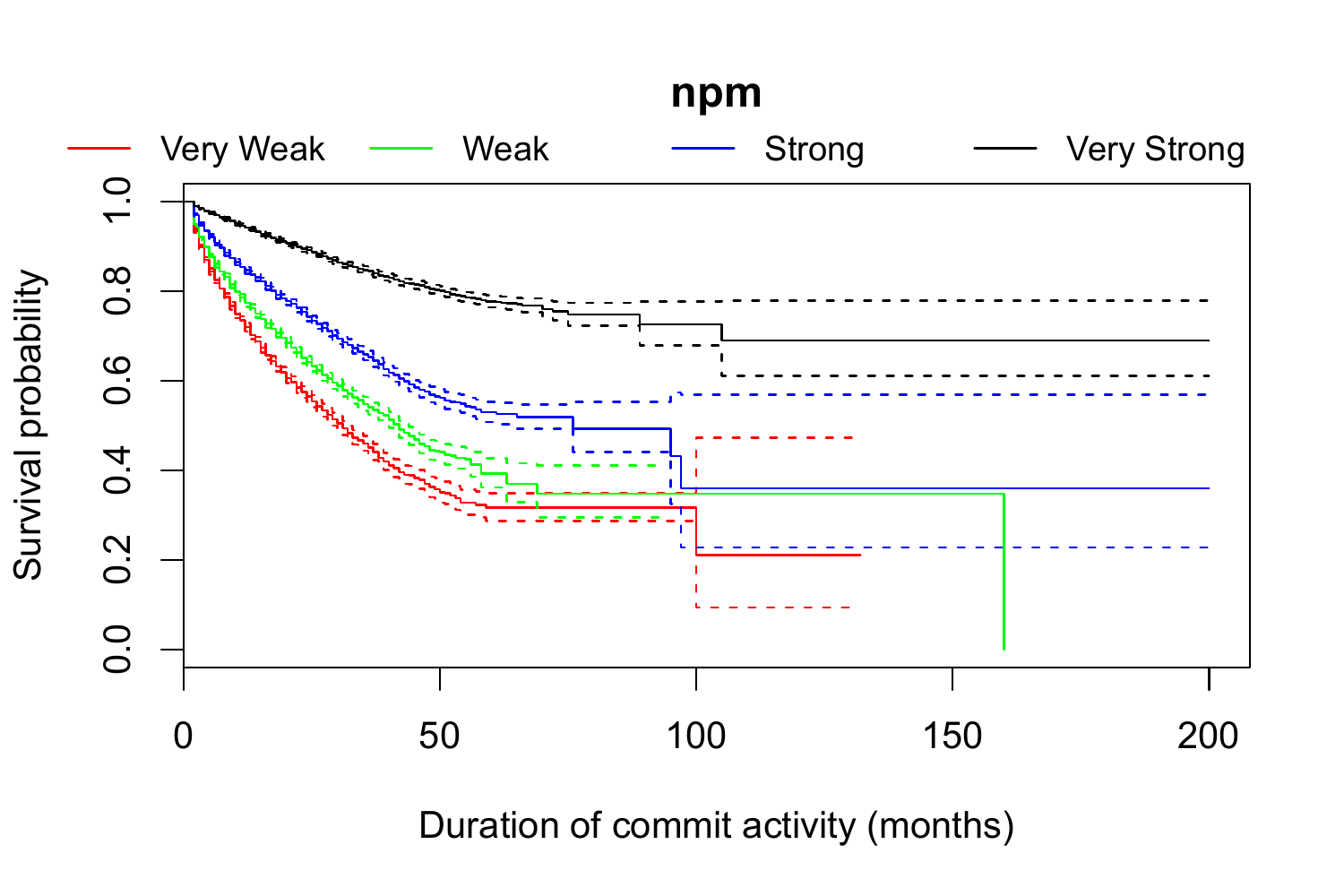}
\end{tabular}
\caption{$H_{1.1}$ -- Survival curves of developers based on communication intensity}
\label{fig:h1_1}
\end{figure}

The survival curves for developers of each communication intensity category are presented in \fig{fig:h1_1}. The survival curves of strong and very strong communication intensity are higher than the ones of weak and very weak categories. Less than 20\% and 40\% of developers with weak and very weak communication intensity are likely to remain in the ecosystem after 10 years for RubyGems and npm respectively, while the survival probability of developers with strong and very strong communication intensity correspond to 25\% and 50\% for RubyGems and 40\% and 70\% for npm respectively.  These results confirm our hypothesis for both ecosystems that \emph{developers who communicate more intensively with other developers in the ecosystem are more likely to remain active for longer periods of time}. 

\subsection*{\textbf{$H_{1.2}$ Developers that communicate less frequently have a higher probability of abandoning the ecosystem sooner.}}

\begin{figure}[!ht]
\centering
\begin{tabular}{c}
\includegraphics[scale=0.49]{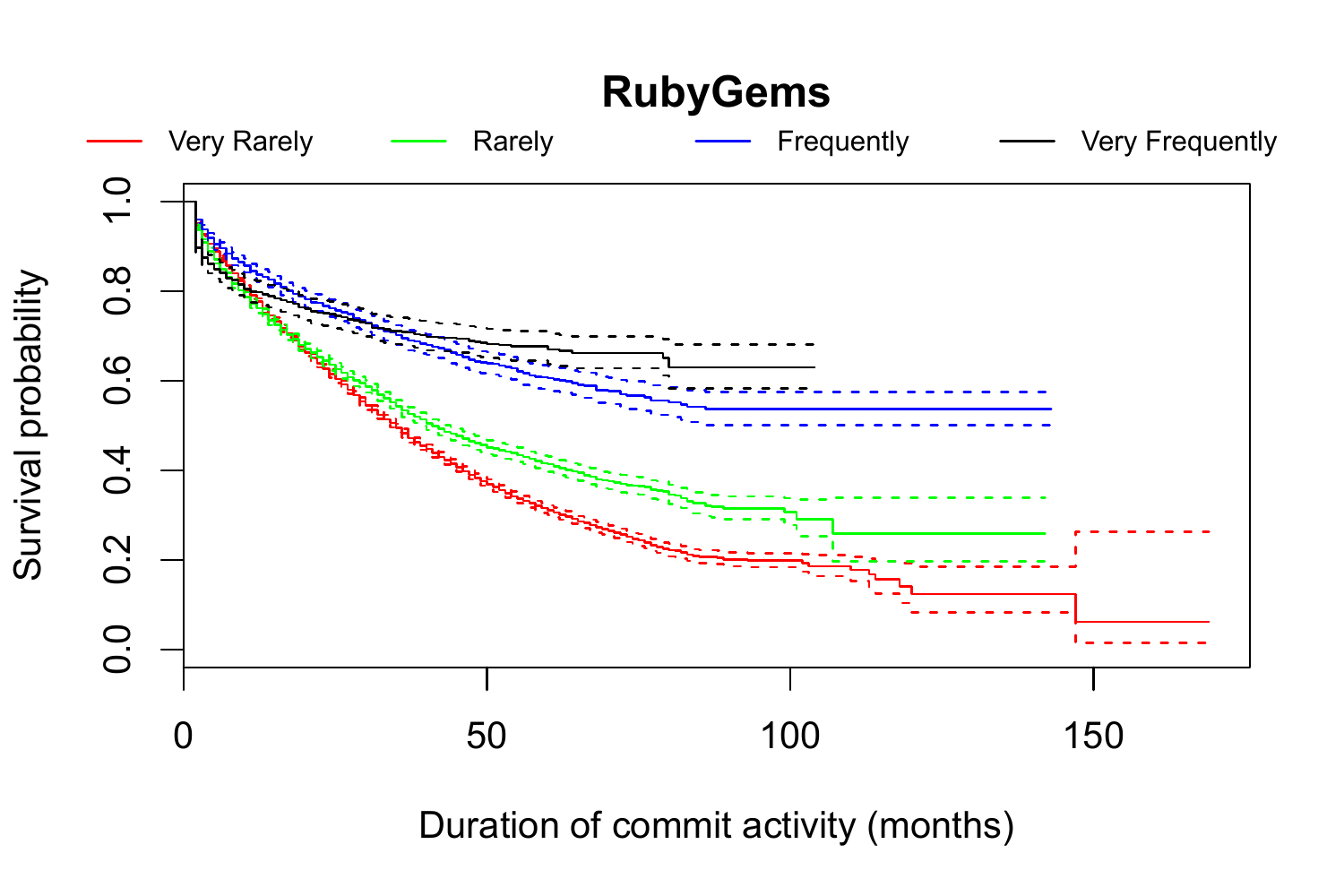}
\\
\includegraphics[scale=0.49]{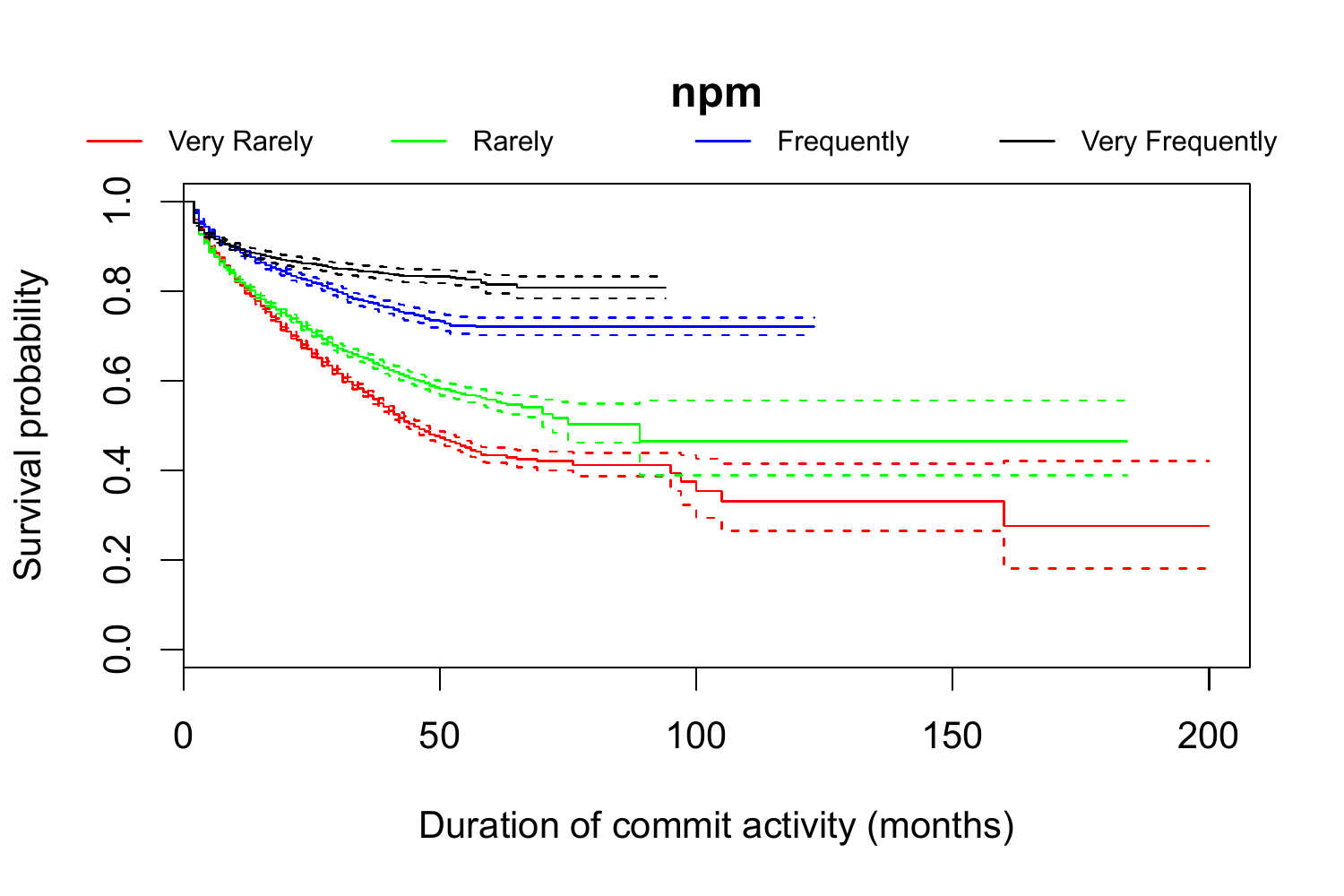}
\end{tabular}
\caption{$H_{1.2}$ -- Survival curves of developers based on communication frequency}
\label{fig:h1_2}
\end{figure}
\fig{fig:h1_2} presents the survival curves with respect to the communication frequency. For both ecosystems, developers who communicate rarely and very rarely have lower survival curves compared to frequently and very frequently communicating developers.
Developers in the very rarely and rarely categories have less than 30\% and 50\% probability of surviving in RubyGems and npm respectively after 100 months. On the contrary, developers in the frequently and very frequently categories have less than 60\% and 80\% probability of remaining active communicators in RubyGems and npm. 
For the RubyGems ecosystem, there was no significant difference (after Bonferroni correction) between the \emph{rarely} and \emph{very rarely} categories. Neither did we find a significant difference between the \emph{frequently} and \emph{very frequently} categories. However, we did find statistically significant evidence for both ecosystems that \emph{developers that communicate rarely or very rarely are more likely to abandon the ecosystem than developers that community frequently or very frequently.}

\subsection*{\textbf{$H_{1.3}$ Developers that do not communicate for a longer period have a higher probability of abandoning the ecosystem sooner.}} 
\begin{figure}[!t]
\centering
\begin{tabular}{c}
\includegraphics[scale=0.49]{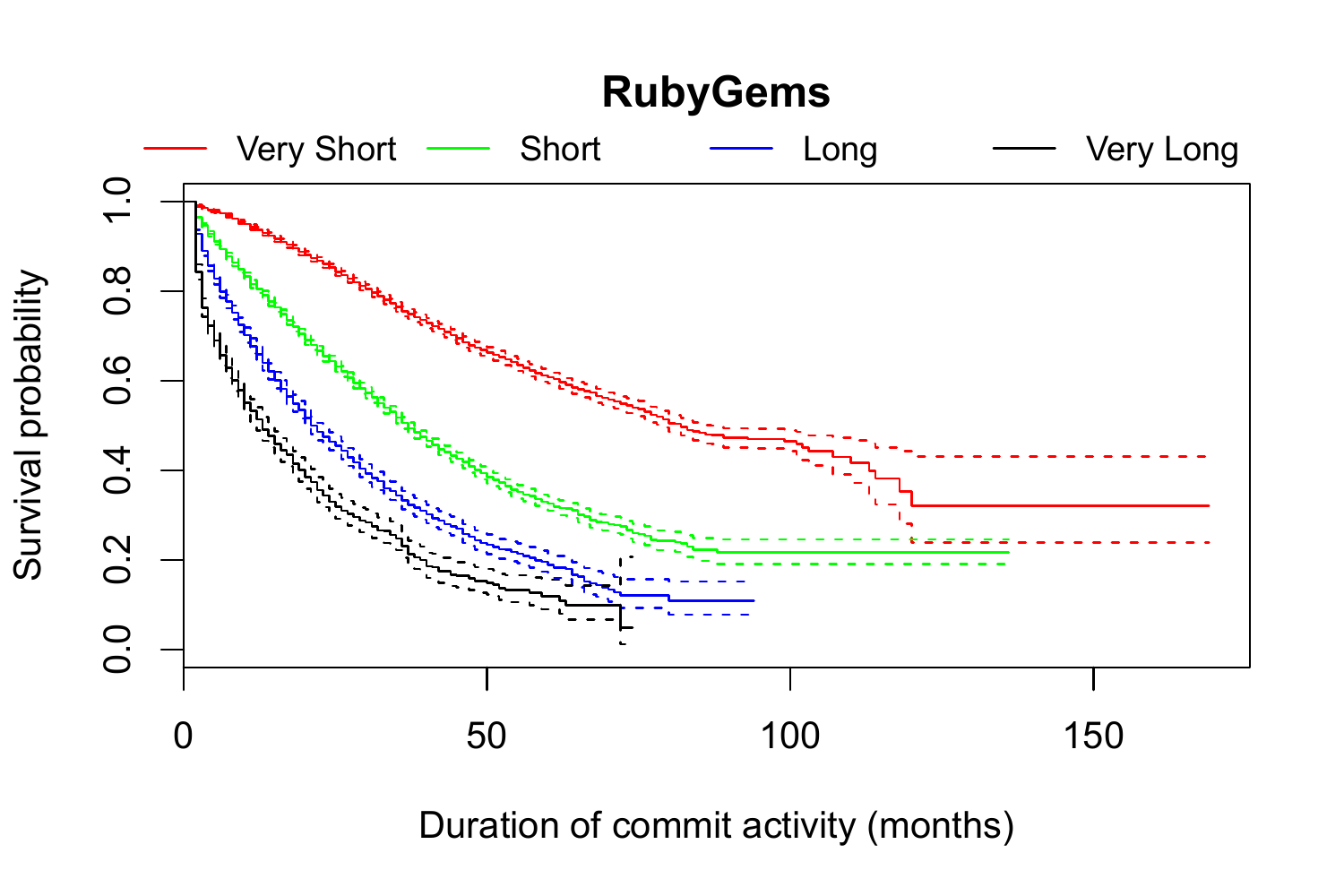}
\\
\includegraphics[scale=0.49]{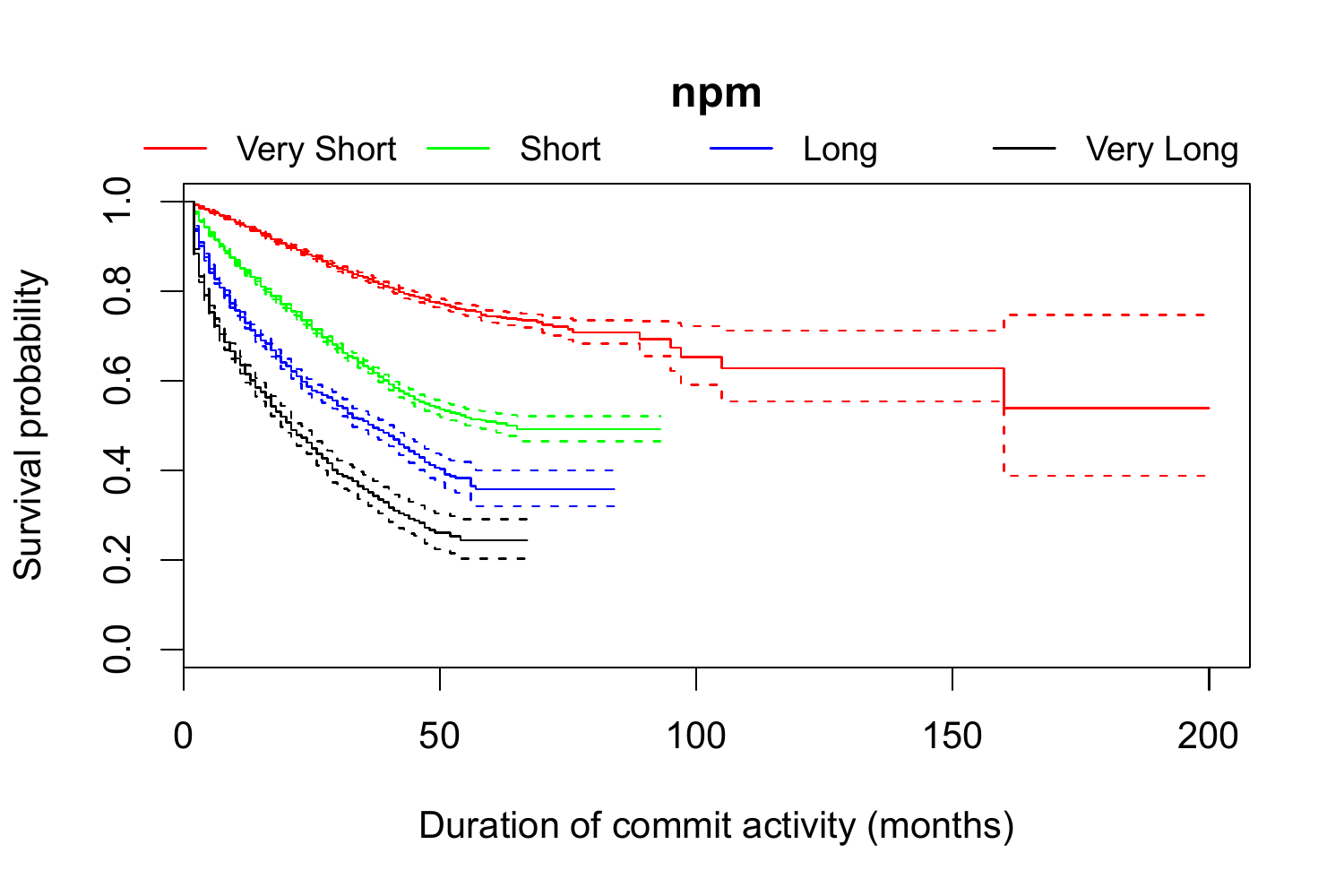}
\end{tabular}
\caption{$H_{1.3}$ -- Survival curves of developers based on social inactivity length}
\label{fig:h1_3}
\end{figure}

The survival curves of each category of communication inactivity are presented in \fig{fig:h1_3}. These curves show that the shorter the period of communication inactivity of a developer, the higher the probability of remaining an active developer in the ecosystem. More concretely, less than 20\% and 40\% of developers with long and very long inactivity periods are likely to remain active in the ecosystem after 50 months for RubyGems and npm respectively. On the contrary, the survival probability of developers with very short and short periods of communication inactivity correspond to less than 70\% and 40\% for RubyGems and less than 80\% and 60\% for npm respectively.
These results confirm our hypothesis that \emph{the longer a developer remains socially inactive, the higher the probability of abandoning the ecosystem}.

\subsection{Survival Analysis of Technical Activity}

\subsection*{\textbf{$H_{2.1}$ Developers  that commit less intensively have a higher probability of abandoning the ecosystem sooner.}}
\begin{figure}[!t]
\centering
\begin{tabular}{c}
\includegraphics[scale=0.49]{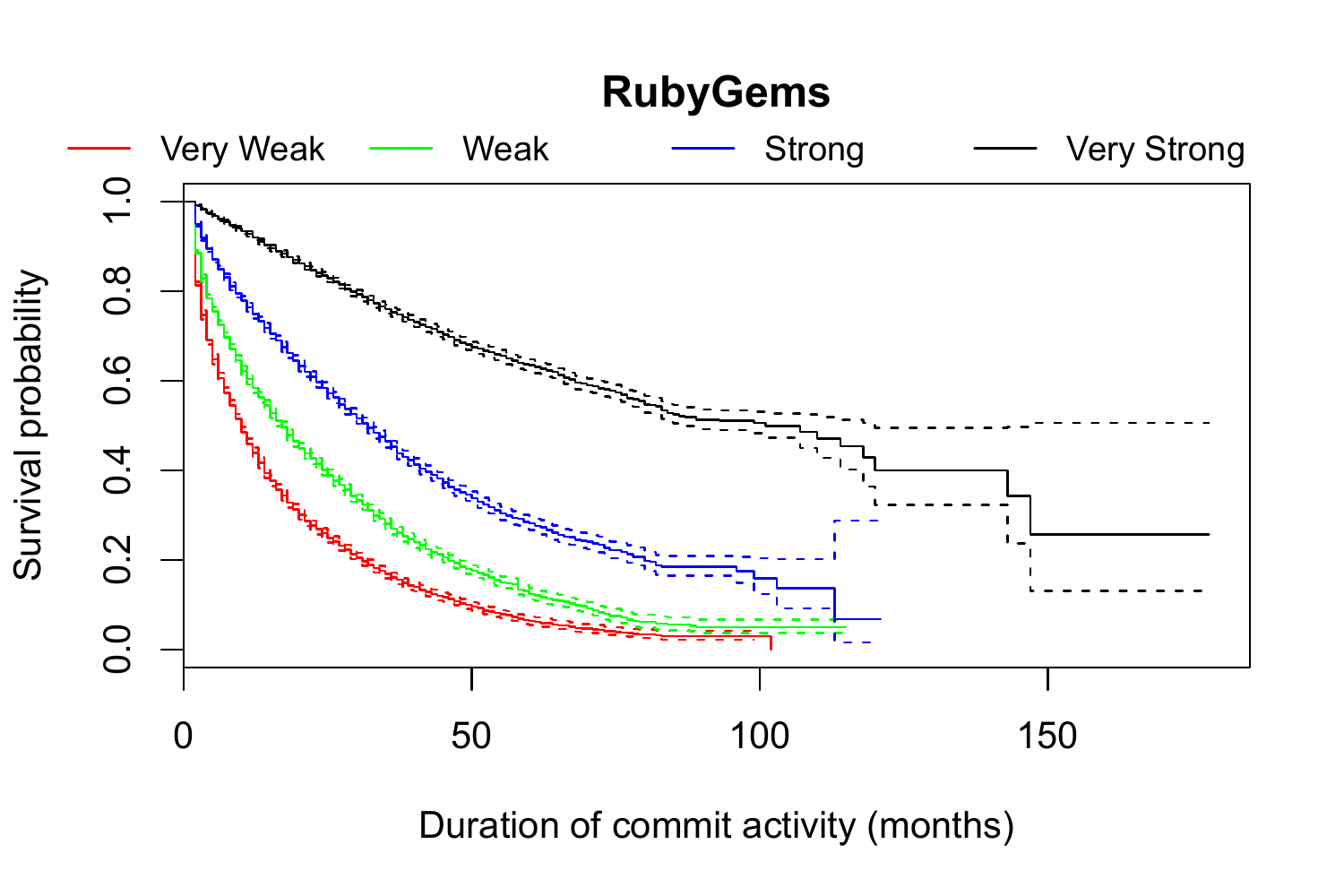}
\\
\includegraphics[scale=0.49]{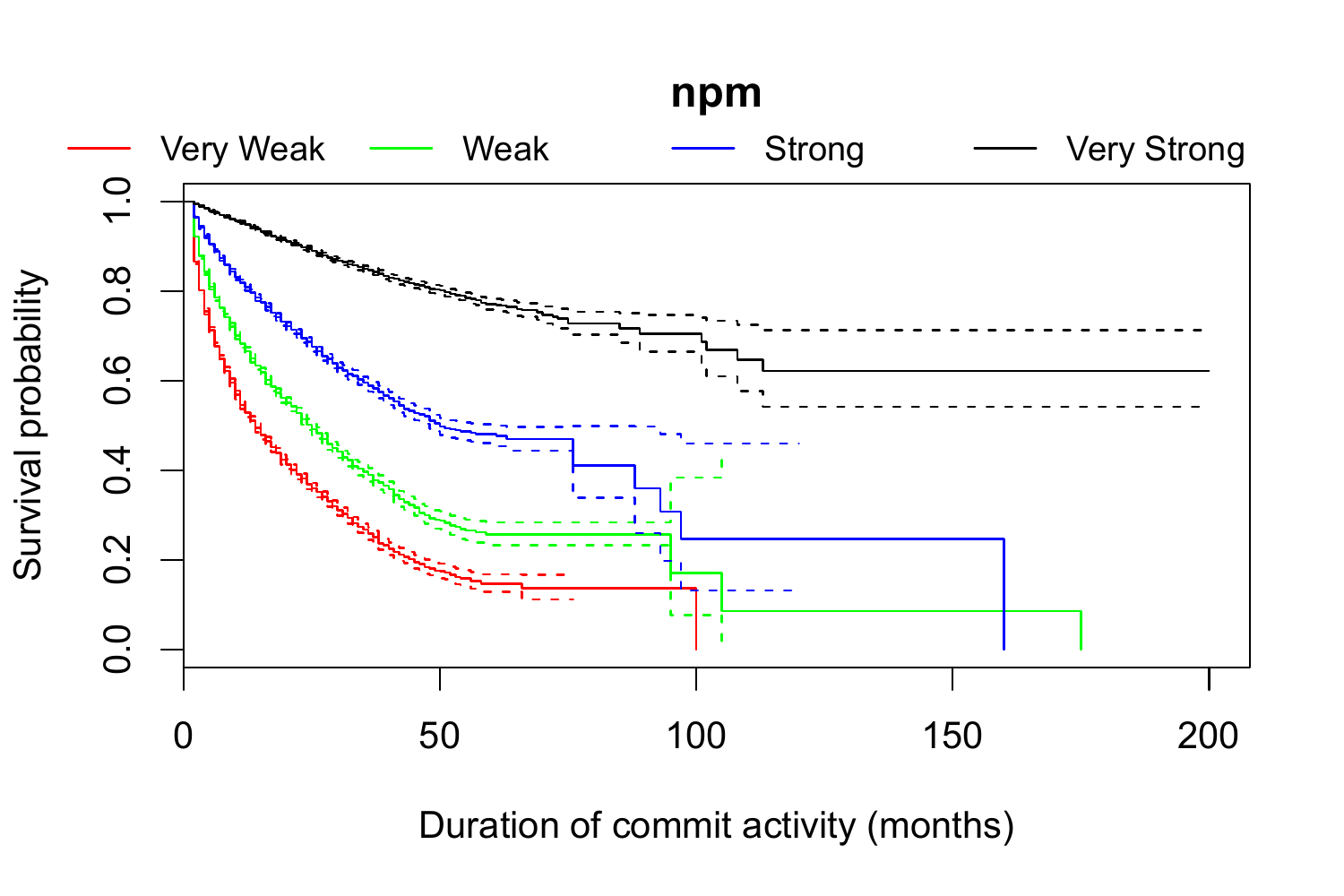}
\end{tabular}
\caption{$H_{2.1}$ -- Survival curves based on commit activity intensity}
\label{fig:h2_1}
\end{figure}

\fig{fig:h2_1} presents the survival curves of commit activity intensity categories for both ecosystems. The survival curves for strong and very strong commit activity intensity are higher than the ones of the two weak intensity categories. After 100 months since joining each ecosystem, less than 10\% and 30\% of developers in the two weak commit activity intensity categories are likely to remain active in RubyGems and npm respectively. The survival probability of developers with strong and very strong commit intensity correspond to 20\% and 60\% for RubyGems and 25\% and 70\% for npm respectively.
This confirms our hypothesis that \emph{the weaker the commit intensity, the higher the probability of abandoning the ecosystem sooner.}

\subsection*{\textbf{$H_{2.2}$ Developers that commit less frequently have a higher probability of abandoning the ecosystem sooner.}}

\begin{figure}[!t]
\centering
\begin{tabular}{c}
\includegraphics[scale=0.49]{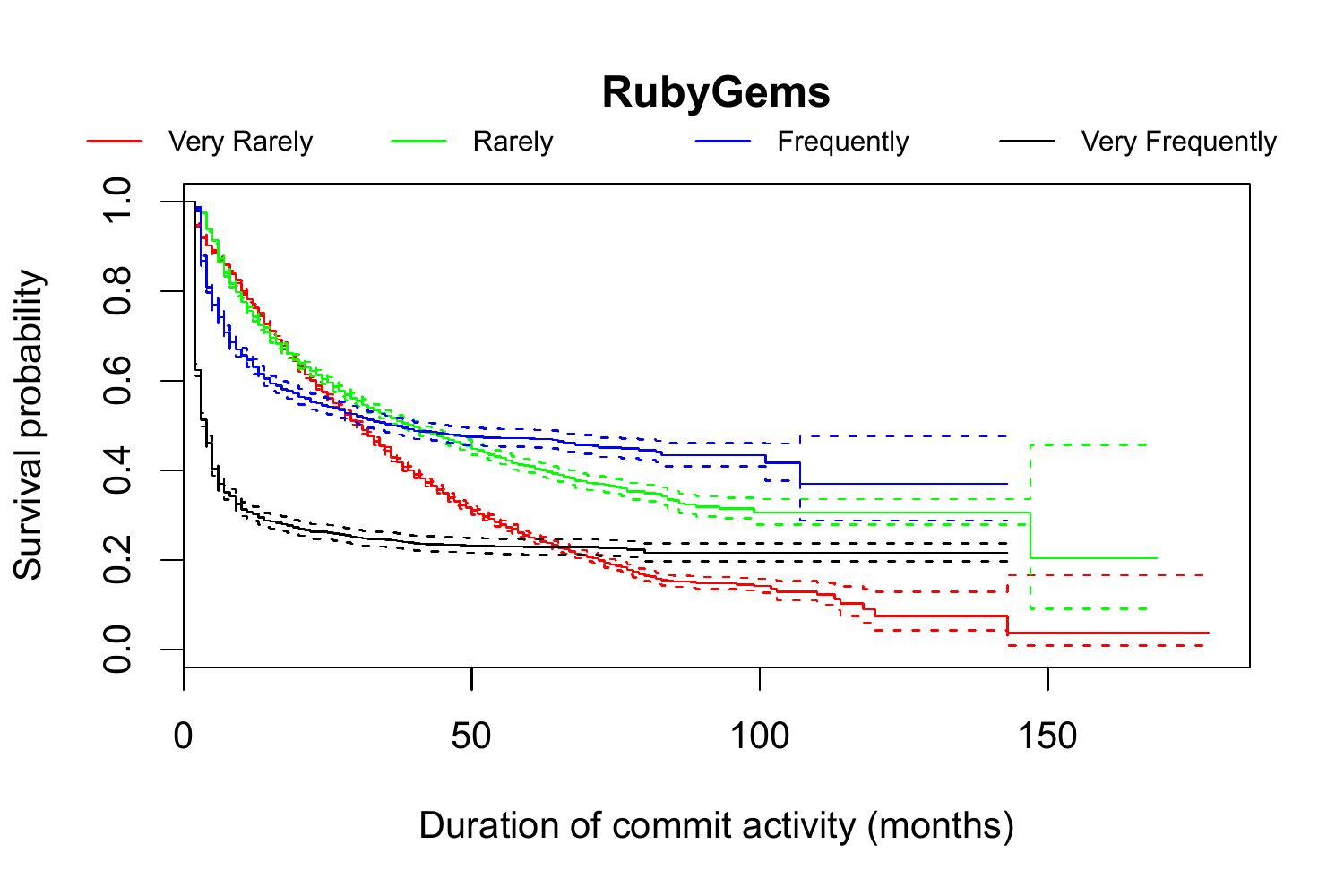}
\\
\includegraphics[scale=0.49]{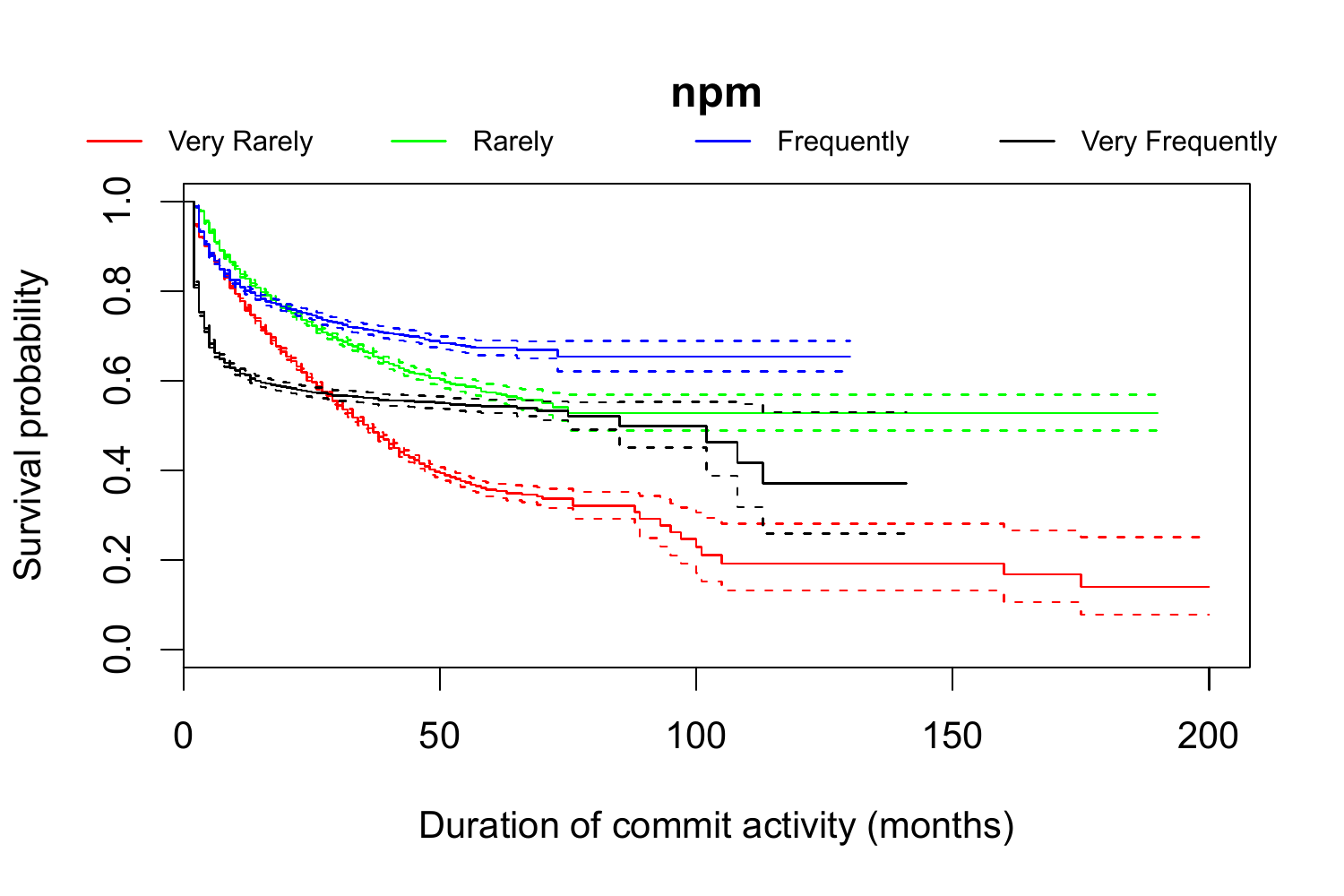}
\end{tabular}
\caption{$H_{2.2}$ -- Survival curves based on commit activity frequency}
\label{fig:h2_2}
\end{figure}
\fig{fig:h2_2} presents the survival curves of the commit frequency categories.
The survival trend differs for RubyGems and npm.
(Very) frequently active developers in RubyGems have lower survival curves for the first months of their activity. More specifically, only 40\% of very frequently active RubyGems developers remain active past the fifth month of their activity, while 50\% of frequently active developers will not abandon the ecosystem after 36 months. In npm, the respective probabilities of remaining active in the ecosystem correspond to 66\% and 72\%.
However, the long term survival probability stabilises over time since for both ecosystems the developers who remain frequently active for more than 20 months have higher survival probabilities.
Concerning the (very) rarely active developers in both ecosystems, their survival probabilities match the ones of frequently active developers during the first months of their activity, but on a longer term the probability to abandon the ecosystem increases (especially for very rarely active developers). 
For the npm ecosystem, not all log-rank tests (after Bonferroni correction) revealed statistically significant differences. However, when regrouping into only two categories of developers (those with frequent or very frequent commit activity on the one hand, and those with rare or very rare commit activity on the other hand), we did find statistically significant differences.
This confirms our research hypothesis that \emph{developers with less frequent commit activity are more likely to abandon the ecosystem}.

\subsection*{\textbf{$H_{2.3}$ Developers that do not commit for longer periods have a higher probability of abandoning the ecosystem sooner.}}

\begin{figure}[!ht]
\centering
\begin{tabular}{c}
\includegraphics[scale=0.49]{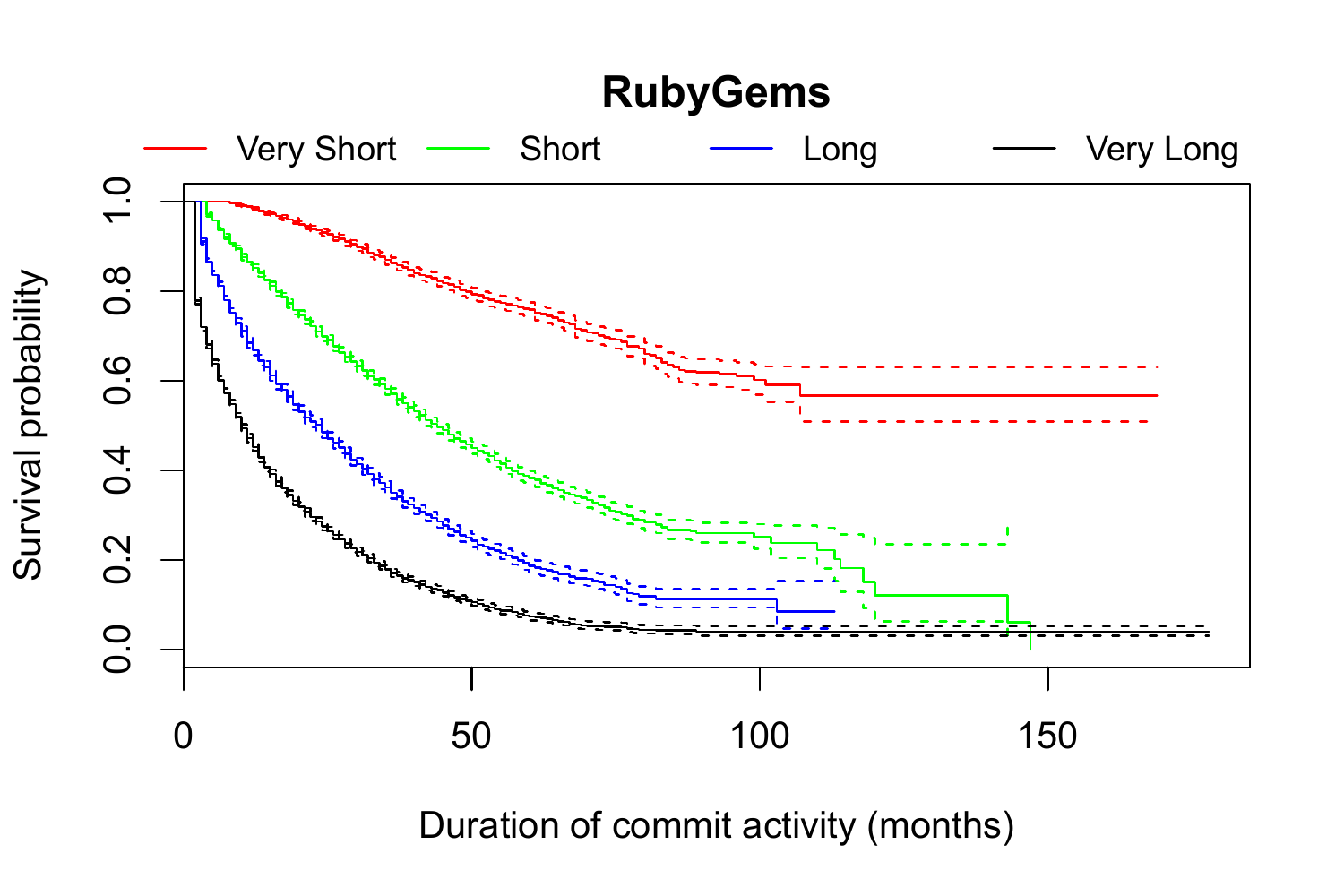}
\\
\includegraphics[scale=0.49]{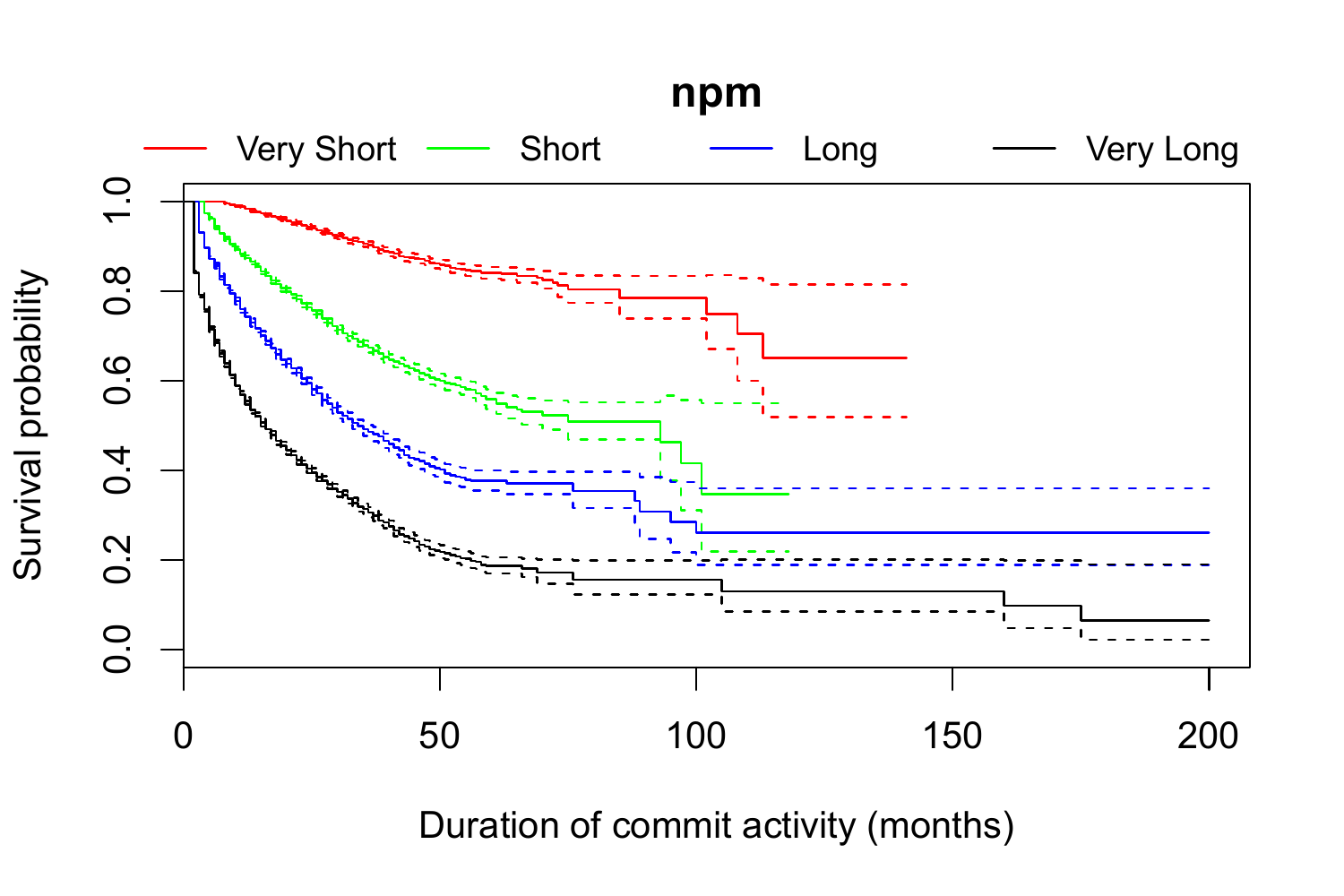}
\end{tabular}
\caption{$H_{2.3}$ -- Survival curves based on commit inactivity length}
\label{fig:h2_3}
\end{figure}

The survival curves of \fig{fig:h2_3} show that the longer a developer remains inactive, the higher the probability of abandoning the ecosystem. More specifically, RubyGems and npm developers that remain inactive for (very) long periods of time have less than 10\% and 25\% probability of remaining active after 100 months since they joined the ecosystem respectively. Developers who remain inactive for very short periods of time have a probability of remaining active close to 60\% and 80\% after 100 months for RubyGems and npm respectively. The survival probability reduces for short inactivity periods, but remains larger than the one of long inactivity categories.	
The data support our hypothesis that \emph{developers who do not commit for longer periods of time have a higher probability of abandoning the ecosystem}.

\section{Core Versus Peripheral Developers}
\label{sec:resultscoreperipheral}

Several empirical studies have investigated the socio-technical structure of open source projects and have shown that the majority of the development effort can typically be attributed to a small percentage of developers~\cite{Mockus:2002}. These developers are referred to as \emph{core} developers and have a substantial and long-term involvement~\cite{Crowston:2006}. On the other hand, \emph{peripheral} developers are contributing more irregularly or their involvement is short-term~\cite{Crowston:2006,Mitchell2017ICSE}.

In this section, we will classify all developers into two categories of \emph{core} and \emph{peripheral} developers within each ecosystem, in order to assess whether the  survival analyses produce different results considering the different contribution frequency and involvement of each type of developer.
A commonly used way to separate core and peripheral developers is to gather their number of commits in the project and compute a threshold at the 80\% percentile~\cite{Crowston:2006,Mockus:2002,Robles:2006,Terceiro2010}. In our work, however, we focus on developer activity that \emph{spanning across multiple projects} within the ecosystem. Project-specific thresholds may introduce bias at an ecosystem level, since a few projects are very intensively developed, e.g., Rails in RubyGems~\cite{Syed2013}.
Because of this, we prefer to use measures that do not depend on a threshold, inspired by the Hirsch index~\cite{Hirsch2005} used to measure productivity and citation impact of scientific publications of a scientific scholar. More specifically, we define two indices to quantify the intensity and spread of a developer's effort in a given ecosystem.

\begin{defn}The Technical Intensity (TI) of a developer $d$ in an ecosystem $E$ is the maximal value $n$ such that $d$ has at least $n$ commits to a project in $E$ to which at least $n$ developers committed. \end{defn}

\begin{defn}The Technical Spread (TS) of a developer $d$ in an ecosystem $E$ is the maximal value $n$ such that $d$ has at least $n$ commits to at least $n$ different projects belonging to $E$. \end{defn}

By considering the $n$ most commits, these indices take into account the skewed distribution of developer contributions within the ecosystem. Thus, \emph{technical intensity TI} considers both the development effort and the project importance within the ecosystem. On the other hand, \emph{technical spread TS} measures both the development effort and the contribution importance to the entire ecosystem.

For each measure, we split contributors into the categories of \emph{core} and \emph{peripheral} based on the median (50th percentile) for \emph{TI} and 65th percentile for \emph{TS}. The reason we chose for the 65th percentile for \emph{TS} is that the 50th percentile equals to the minimum value of \emph{TS=1} since a large portion of the developer population does not contribute intensively to many projects.
The range of values for $TI$ and $TS$ corresponding to both categories of contibutors are summarised in \tab{tab:core_per_stats}.

\begin{table*}[!ht]
\renewcommand{\arraystretch}{1.3}
\caption{Descriptive statistics of Technical Intensity and Technical Spread of ecosystem contributors}
\label{tab:core_per_stats}
\centering
\begin{tabular}{c|r|c|c|c}
Index & Contributor & Binning & RubyGems & npm\\
 & category & (percentile) &  & \\
\hline
Technical Intensity & core & $>50$th & $TI \in [3,578]$ & $TI \in [3,1034]$\\
($TI$) & peripheral&  $<50$th& $TI \in [1,2]$ & $TI \in [1,2]$\\
\hline
Technical Spread & core & $>65$th & $TS \in [2,24]$ & $TS \in [2,39]$\\
($TS$) & peripheral & $<65$th& $TS=1$ & $TS=1$ 
\end{tabular}
\end{table*}

We replicated our survival analyses and log-rank tests (with Bonferroni correction) for all research hypotheses, considering the subgroups of \emph{core} and \emph{peripheral} developers based on \emph{TI} and \emph{TS}, respectively.
Since most of our research hypotheses were confirmed for the different developer categories, we will only present those results that differ with respect to the previously reported results on the entire developer population of each ecosystem.

For $H_{1.1}$, when considering core developers based on \emph{TS} in the npm ecosystem, no statistical difference was found between the communication intensity categories. This suggests that \emph{npm core developers that work on multiple projects have a similar probability of remaining active in the ecosystem regardless of their communication intensity}. The same results occurred for the peripheral developers based on \emph{TI}, suggesting that \emph{npm peripheral developers with a low number of commits to small projects have a similar probability of remaining active regardless of their communication intensity.}

\begin{figure}[!ht]
\centering
\includegraphics[width=\columnwidth]{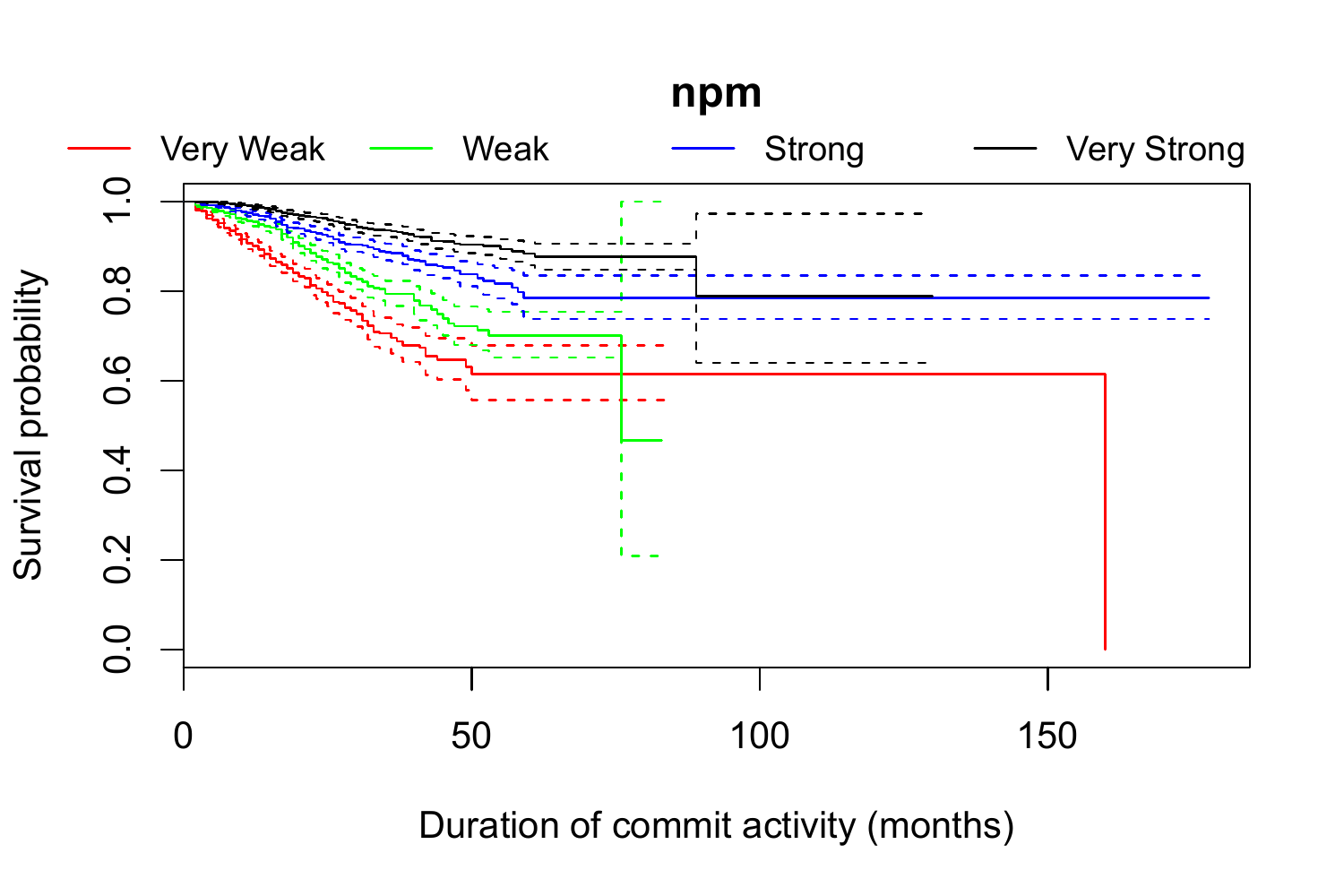} 
(a) Core developers based on \emph{TS}\\
\includegraphics[width=\columnwidth]{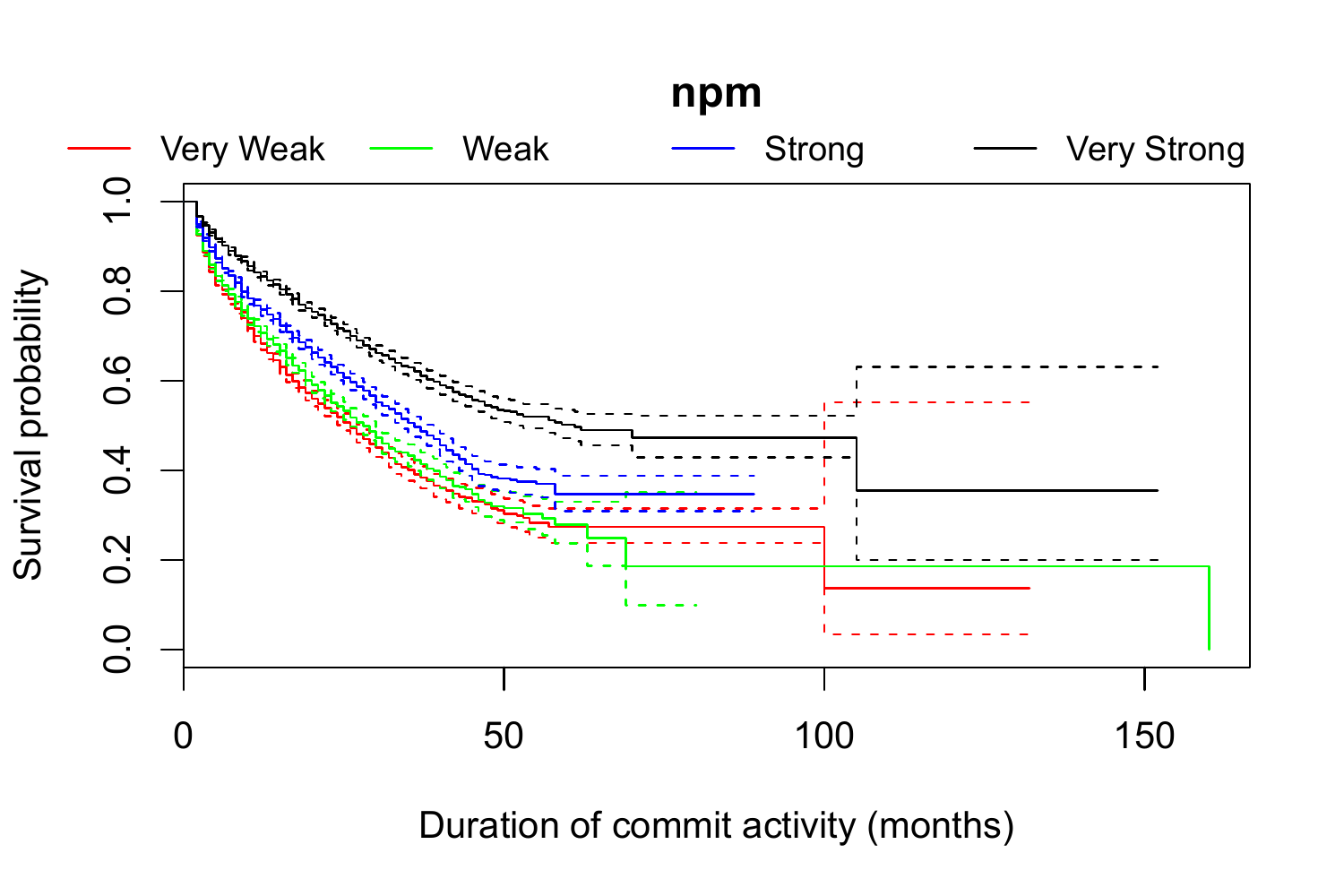}
(b) Peripheral developers based on \emph{TI}
\caption{$H_{1.1}$ -- Survival curves for \textbf{npm} based on communication intensity.}
\label{fig:h_1_1_core_peripheral}
\end{figure}

\fig{fig:h_1_1_core_peripheral} presents the results for both contributor categories of the npm ecosystem. Compared to the results of the entire population (see \fig{fig:h1_1}) we observe that, although peripheral \emph{TI} developers with very strong communication intensity have a higher probability of remaining active in each ecosystem compared to the other intensity categories of peripheral developers, they have a lower probability of remaining active when compared to core developers. In turn, although core \emph{TS} developers have a similar probability of remaining active in the ecosystem regardless of their communication intensity, the probability of core developers remaining active is larger compared to peripheral developers.

\begin{figure}[!ht]
\centering
\includegraphics[width=\columnwidth]{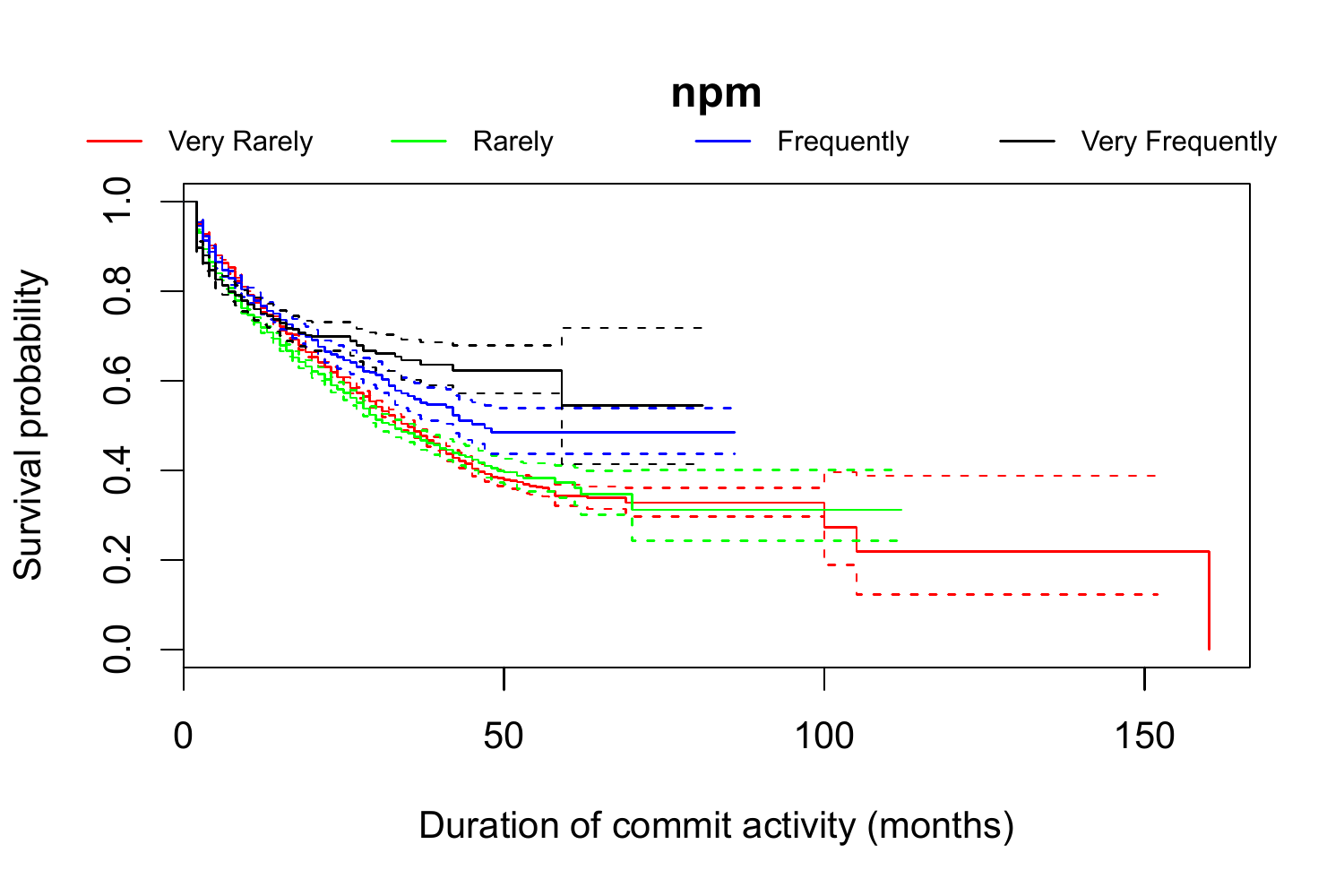}
(a) Peripheral developers based on \emph{TI} \\
\includegraphics[width=\columnwidth]{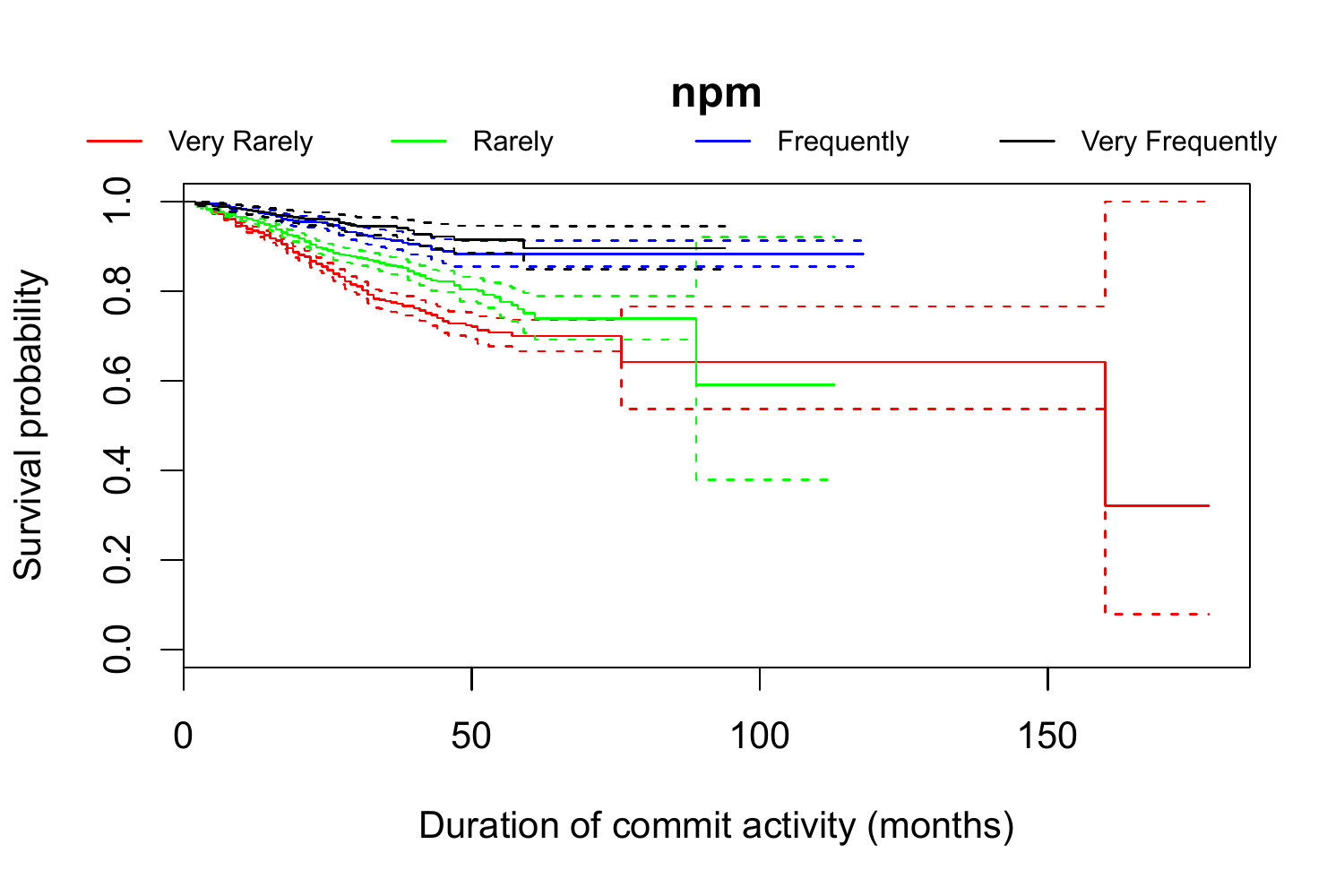}
 (b) Core developers based on \emph{TS}
\caption{$H_{1.2}$ -- Survival curves for \textbf{npm} based on communication frequency.}
\label{fig:h_1_2_core_peripheral}
\end{figure}

Concerning $H_{1.2}$, we also observed some differences in the behaviour between core and peripheral developers, but only for npm. 
\fig{fig:h_1_2_core_peripheral} reveals that, compared to the results for the entire population of npm in \fig{fig:h1_2}, the survival probabilities for peripheral \emph{TI} developers decrease, suggesting that \emph{peripheral developers with a low number of commits to small projects are less likely to remain active in the ecosystem regardless of the frequency of their communication activity}. In contrast, the survival probabilities for core \emph{TS} developers suggest that \emph{core developers who contribute to many projects in the ecosystem are more likely to remain active in the ecosystem regardless of their communication frequency.}

\begin{figure}[!ht]
\centering
\includegraphics[width=\columnwidth]{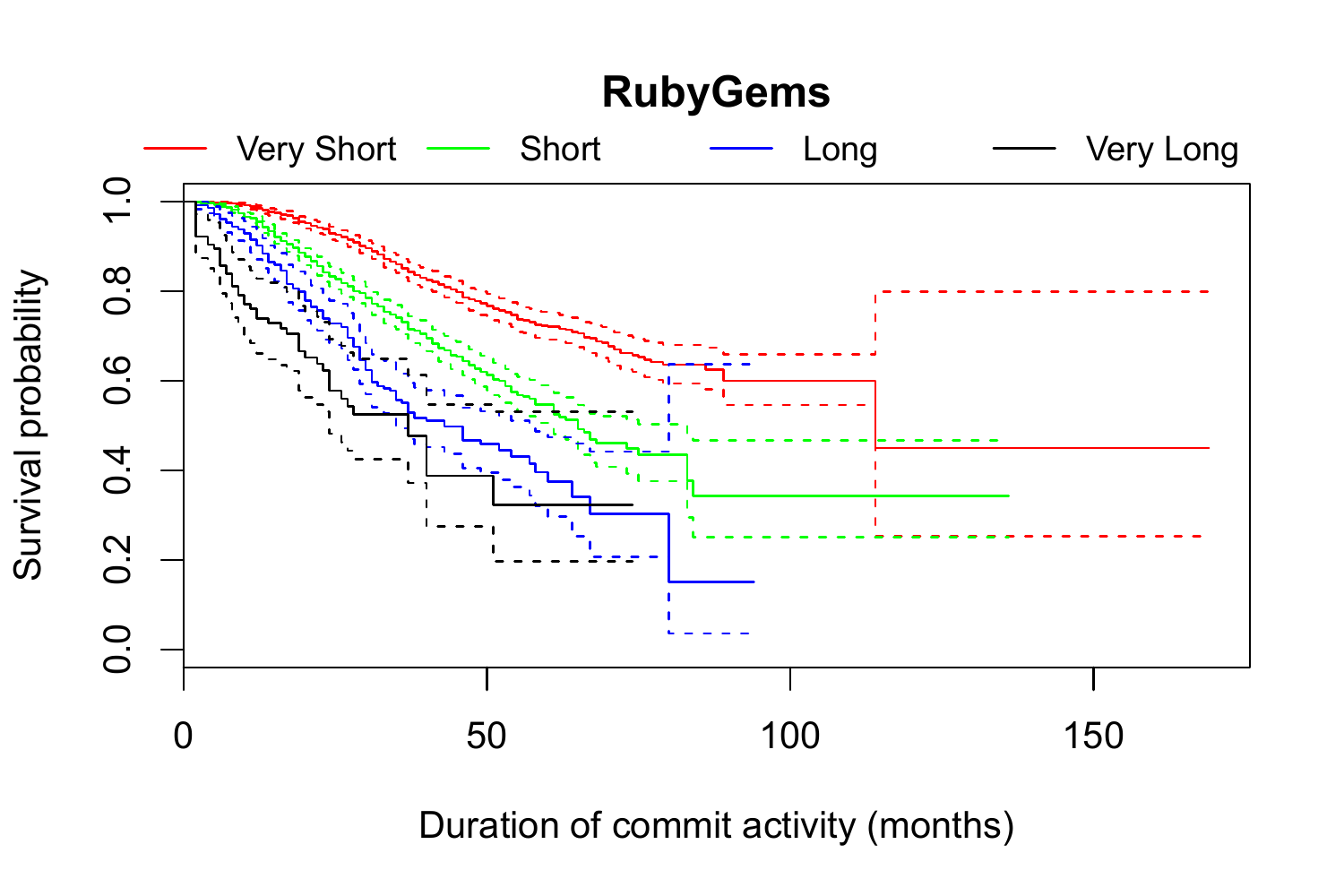}
\includegraphics[width=\columnwidth]{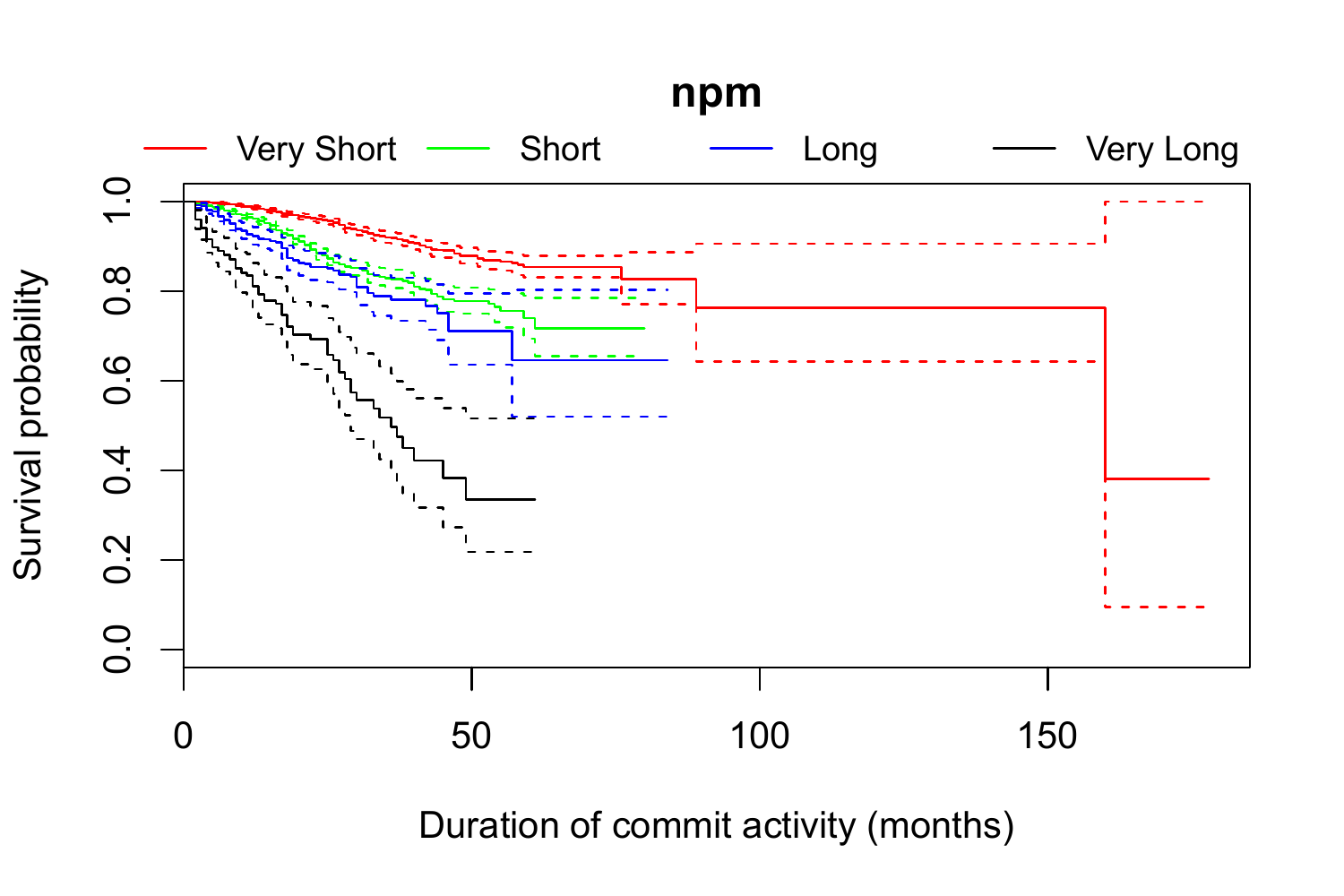}
\caption{$H_{1.3}$ -- Survival curves based on social inactivity length for core developers based on \emph{TS}.}
\label{fig:h_1_3_core_peripheral}
\end{figure}

Concerning $H_{1.3}$, we found differences for core \emph{TS} developers in both ecosystems and present these results in Figure~\ref{fig:h_1_3_core_peripheral}. We did not find any statistical evidence that core developers who contribute to many projects in the ecosystem have significant differences in their probability of remaining active for longer periods of time depending on the length category of their social inactivity period.

Finally, we found some differences concerning $H_{2.2}$ for both ecosystems.
More specifically, for npm we found that both \emph{core \emph{TI} developers with a large number of commits to large projects and peripheral \emph{TI} developers with a low number of commits to small projects are more likely to abandon the ecosystem when they commit less frequently.}
For RubyGems, however, \emph{all groups of core \emph{TI} developers have a similar probability of abandoning the ecosystem regardless of their commit frequency.} Additionally, we found that both core and peripheral \emph{TS} developers in RubyGems have a similar probability of abandoning the ecosystem regardless of their commit frequency. These results show that the effect of commit frequency on developer abandonment differs among the two ecosystems: in npm frequent commit activity is indicative of a prolonged contributor presence in the ecosystem for both core and peripheral \emph{TI} developers, while in RubyGems it does not appear to be a useful factor for distinguishing between core and peripheral contributors who are more likely to remain active in RubyGems for longer periods of time.

\section{Discussion and Limitations}
\label{sec:discussion}
Whether and how a software ecosystem successfully evolves over time depends to a large extent on the activeness and interaction of its developer community. Developer turnover and abandonment pose important threats with respect to knowledge loss~\cite{IzquierdoCortazar2009,Rigby:2016}, and resources and time to familiarise new members~\cite{Fritz:2010}. 
Such problems may have an important impact in a software ecosystem because of the many interdependencies between software projects and the risk of project abandonment affecting its transitive dependents. Therefore, it is of great importance to determine factors affecting developer retention, which can be used to predict risks associated with developer abandonment, and to mitigate such risks early. Our empirical study is a first step in this direction, by quantifying developer risk/retention based on information stored in software repositories, thus facilitating the automation of such approaches.

A limitation of our current work is that we consider as developer any contributor that actively commits changes to the project's source code repository, even though these changes may not necessarily involve source code files. 
In future work we will refine this definition, by taking into account the types of files touched by each contributor, similar to the work of Vasilescu \etal \cite{Vasilescu2014}.

The analysis presented in Section~\ref{sec:resultscoreperipheral} reveals that developer retention can be improved when contributors are active frequently and have strong contribution intensity, regardless of whether they are core or peripheral developers. However, there is not necessarily a causal relation between the factors we identified and the actual reason why developers abandon software ecosystems. Therefore, our quantitative analyses do not allow us to claim that developers abandon the ecosystem \emph{because} they do not engage in discussions. However, we can treat the factors that affect developer retention as \emph{symptoms} that can be used to predict the probability of developers abandoning the ecosystem.

Our analyses were based on the assumption that developers are considered to have abandoned the ecosystem if they have not been committing to a project's repository for at least one year. Other studies use different thresholds to find abandoners. For example, Lin et al.~\cite{Lin:2017} use a threshold of only 180 days to determine whether a developer has abandoned a project. They also used different thresholds (30 and 90 days) and found similar trends in the survival curves. However, they focused on open source projects in which a minority of developers were volunteers, since the considered projects were supported by software companies for the majority of tasks. In contrast, our study focuses on two ecosystems with a large portion of volunteers. Therefore, using similar thresholds as those proposed by Lin \etal~\cite{Lin:2017} would likely exclude many peripheral developers who contribute consistently, but might remain inactive for more than six months between active periods of time.

Our empirical study is inherently limited by the factors we have taken into account for affecting developer retention. 
We could extend this list with many other possible factors, such as developer workload, expertise, seniority, gender, activity type, etc. Including such factors would require a different and more fine-grained analysis that focuses on developer characteristics, rather than ecosystem activity factors.

Another limitation of our work lies in the fact that we examined different factors of developer retention independently. In our future work, we will combine different factors in our survival analysis to gain a better understanding of developer profiles that are more likely to remain active longer in an ecosystem.

\section{Threats to Validity}
\label{sec:threats_to_validity}

\subsection{Internal Validity}
We considered factors concerning social and technical activity of developers in software ecosystems, as well as their intensity and frequency. However, additional external factors may influence a developer's decision to abandon an ecosystem, including factors such as personal issues that cannot be quantified or predicted.

\subsection{Construct Validity}
Our study only considered developer communication information extracted from the GitHub commenting mechanisms. However, developers might use different platforms to communicate such as mailing lists. To mitigate this risk, we explored one mailing list of npm\footnote{\url{https://groups.google.com/forum/#!forum/npm-}} and two of RubyGems\footnote{\url{https://groups.google.com/forum/#!forum/rubygems-org}},\footnote{\url{https://groups.google.com/forum/#!forum/rubygems-developers}}. However, a requirement to using this information is to match developer identities between the mailing list participants and GitHub developers, which can pose additional threats to our work. Considering that the activity in these mailing lists was infrequent compared to the activity in GitHub comments, and considering that a small portion of mailing list discussions addresses implementation details~\cite{Guzzi:2013}, we omitted this data source from our study.
As another possible threat, we did not merge developer identities and thus, different GitHub accounts might correspond to the same developer. However, this threat is limited considering that using multiple accounts in GitHub is not very common~\cite{Vasilescu:2015MSR}.

Due to the experimental setup of our study, our dataset consists of a considerable subset of packages of each ecosystem ($>$56\% according to Table~\ref{tab:dataset_statistics}). However, not all ecosystem packages are hosted in GitHub even though some of them may have explicit dependencies to or from the packages we considered. Although a recent study reports that the majority of JavaScript and Ruby ecosystems are hosted on GitHub~\cite{Kikas:2017}, there is a threat that developers that we have categorised as abandoners might actually remain active in other ecosystem projects not hosted on GitHub. 

\subsection{External Validity}
We have only analysed the socio-technical activity of two different open source ecosystems. Although this included over 30k and 60k developers for each ecosystem, we cannot generalise our results to other software ecosystems.
In particular, the phenomenon of developer abandonment might be quite different for smaller and less popular ecosystems. Also, the results may be different for mixed source or proprietary (inner source) software ecosystems.

\section{Conclusion}
\label{sec:conclusion}
In this article, we performed an extensive empirical study over two large, long-lived software ecosystems: RubyGems and npm. We examined the relationship between the frequency and intensity of socio-technical activity and developer retention. Social activity was measured in terms of communications involving multiple developers through GitHub comments, while technical activity was measured through commits.

Our findings show that developers have a higher probability of abandoning an ecosystem when they: (1) do not communicate with other developers; (2) do not have a very strong social and technical activity intensity; (3) communicate or commit less frequently; and (4)  do not communicate or commit for a longer period of time.

Additionally, we found some notable differences in developer retention when distinguishing between core and peripheral developers of each ecosystem. For example, the frequency of contributions of core developers do not seem to affect the longevity of their contributions.

We also found differences between the two ecosystems in which factors are likely to impact developer retention. For example, in RubyGems, the factor of social communication frequency does not seem useful to predict developer abandonment. For npm, on the other hand, it is the factor of commit frequency that does not seem useful to predict developer abandonment.

Such observations could be used to build ecosystem-dependent models to automate the prediction of developers with a high probability of abandoning the ecosystem and, as such, reduce the risks associated to knowledge loss. 

\begin{acknowledgements}
This research was carried out in the context of FNRS cr\'edit de recherche J.0023.16 entitled ``Analysis of Software Project Survival'' and the bilateral collaborative research program FRQ-FNRS 30440672 entitled ``Towards an Interdisciplinary Socio-Technical Methodology and Analysis of Software Ecosystem Health''.
\end{acknowledgements}



\end{document}